\documentclass[11pt,a4paper]{article}
\usepackage[margin=2cm]{geometry}
\usepackage[T1]{fontenc}
\usepackage[utf8]{inputenc}
\usepackage[noadjust]{cite}
\usepackage{lmodern}
\usepackage{amsthm}
\usepackage{thmtools}
\makeatletter
\@ifundefined{newcounteralias}{}{%
  \renewcommand\thmt@autorefsetup{%
    \@xa\def\csname\thmt@envname autorefname\@xa\endcsname
      \@xa{\thmt@thmname}%
  }%
}
\makeatother
\newcommand{\cost}{\mathsf{cost}}

\usepackage{thm-autoref}
\usepackage[shortlabels]{enumitem}
\usepackage[center]{caption}
\usepackage{latexsym}
\usepackage{amsfonts}
\usepackage{graphicx}
\usepackage{mathrsfs}
\usepackage[normalem]{ulem}
\usepackage{amssymb, mathdots, mathtools, xkeyval, tikz, tkz-euclide, adjustbox, float, array, subcaption, listings}
\usepackage{comment}
\usepackage{xspace}
\usepackage[colorlinks=true,
linkcolor =blue,
citecolor=blue,
urlcolor=blue]{hyperref}
\usepackage[capitalise,noabbrev]{cleveref}

\theoremstyle{plain}

\newtheorem{theorem}{Theorem}[section]
\newtheorem{lemma}[theorem]{Lemma}

\newtheorem{conjecture}[theorem]{Conjecture}
\newtheorem{question}[theorem]{Question}

\newtheorem{claim}{Claim}[theorem]
\newenvironment{claimproof}{\noindent\textit{Proof of Claim \theclaim:}}{\hfill$\lrcorner$\par\smallskip} % To remove numbering, delete '\theclaim'

\crefname{claim}{Claim}{Claims}
\crefname{conjecture}{Conjecture}{Conjectures}
\Crefname{conjecture}{Conjecture}{Conjectures}
\crefname{question}{Question}{Questions}
\Crefname{question}{Question}{Questions}

\newcommand{\MIS}{\textsc{MIS}\xspace}
\newcommand{\MWIS}{\textsc{MWIS}\xspace}
\newcommand{\col}[1]{\textsc{#1-Coloring}\xspace}
\newcommand{\lcol}[1]{\textsc{List #1-Coloring}\xspace}

\newcommand{\NP}{\textsf{NP}\xspace}
\newcommand{\QP}{\textsf{QP}\xspace}
\newcommand{\Oh}{\mathcal{O}}
\newcommand{\subKfour}{K_4^{\vcenter{\hbox{\scalebox{0.65}{$\scriptstyle\bullet$}}}}}
\newcommand{\mc}{\mathsf{max\text{-}cut}}

\renewcommand{\phi}{\varphi}

\newcommand{\cB}{\mathcal{B}}
\newcommand{\cC}{\mathcal{C}}

\newcommand{\cT}{\mathcal{T}}

\usepackage{todonotes}

\title{What remains of the planar induced minor conjectures?%
\thanks{The use of AI tools in research was partially supported by grant of access to OpenAI models through the ChatGPT for Academic Researchers program
and via Claude Team plan for research teams.}}

\author{
    Jadwiga Czyżewska\thanks{University of Warsaw, Poland (\textsf{j.czyzewska@mimuw.edu.pl}).	
		Supported by Polish National Science Centre SONATA BIS-12 grant number 2022/46/E/ST6/00143.}
	\and Tomáš Masařík\thanks{University of Warsaw, Poland (\textsf{masarik@mimuw.edu.pl}).
			Supported by the Polish National Science Centre SONATA-17 grant number 2021/43/D/ST6/03312.}
	\and Marcin Pilipczuk\thanks{University of Warsaw, Poland (\textsf{m.pilipczuk@mimuw.edu.pl}).
			Supported by Polish National Science Centre SONATA BIS-12 grant number 2022/46/E/ST6/00143.}
	\and Paweł Rzążewski\thanks{Warsaw University of Technology, Poland (\textsf{pawel.rzazewski@pw.edu.pl}). Supported by the National Science Centre grant 2024/54/E/ST6/00094.}}
\date{}

\hypersetup{hypertexnames=false}
\begin{document}
	\begin{titlepage}
	\date{}
	\maketitle

\begin{abstract}
    Dallard, Milani\v{c}, and \v{S}torgel, and independently Gartland and Lokshtanov, conjectured that for every planar graph $H$, \textsc{Maximum Weight Independent Set} and \textsc{List 3-Coloring} are polynomial-time solvable in $H$-induced-minor-free graphs.
    As a structural explanation, Gartland and Lokshtanov further conjectured that such graphs admit balanced separators dominated by a constant number of vertices.
    Bonnet and Chang recently refuted the independent-set part of the algorithmic conjecture, by showing that \textsc{Maximum Independent Set} is \textsf{NP}-hard in graphs excluding the $5\times5$ grid as an induced minor.

    We revisit both conjectures.
    We show that the graphs used by Bonnet and Chang exclude much simpler planar graphs, including several $16$-vertex trees, and we introduce a second, independent family of hard instances.
    We also prove that \textsc{3-Coloring} is \textsf{NP}-hard in both settings, thus refuting the algorithmic conjecture in full, even in its quasipolynomial version, unless $\textsf{NP}\subseteq\textsf{QP}$.
    On the structural side, we identify planar graphs $H$ for which the structural conjecture holds, while the algorithmic one fails.
    Thus, constantly dominated balanced separators cannot by themselves explain the tractability of \textsc{Maximum Weight Independent Set} and \textsc{List 3-Coloring}.
\end{abstract}

	\thispagestyle{empty}

\tableofcontents
\end{titlepage}
\thispagestyle{empty}
\newpage\setcounter{page}{1}

\section{Introduction}
Graph minor theory is one of the cornerstones of structural graph theory.
A graph $H$ is a \emph{minor} of a graph $G$ if $H$ can be obtained from $G$ by deleting vertices and edges and by contracting edges.
A remarkable feature of this theory is that excluding \emph{any} fixed graph as a minor already imposes strong and uniform structure.
In particular, for every fixed graph $H$, the class of $H$-minor-free graphs has bounded average degree~\cite{DBLP:books/daglib/0030491} and admits balanced separators of size $\Oh_H(\sqrt n)$~\cite{AlonSeymourThomas1990, BalancedSeparatorLinearTime}. 
As observed by Alon, Seymour, and Thomas, the latter yields a polynomial-time approximation scheme (PTAS) for \textsc{Maximum Independent Set} (\MIS) in $H$-minor-free graphs, extending the classic approach of Lipton and Tarjan for planar graphs~\cite{LiptonTarjan1980}.
Moreover, by the Graph Minor Theorem of Robertson and Seymour~\cite{RobertsonSeymour2004}, every minor-closed class is characterized by a finite set of excluded minors.

Within this theory, planar graphs mark a sharp boundary.
By the Grid Minor Theorem~\cite{RobertsonSeymour1986}, for every planar graph $H$, the class of $H$-minor-free graphs has bounded treewidth, and hence admits balanced separators of bounded size.
Conversely, if $H$ is not planar, then all planar graphs, including grids of arbitrarily large treewidth, are $H$-minor-free.
Bounded treewidth has far-reaching algorithmic consequences.
By Courcelle's theorem and its optimization variants~\cite{Courcelle1990,ArnborgLagergrenSeese1991}, many problems that are \NP-hard in general, such as \textsc{Maximum Weight Independent Set} (\MWIS) and \lcol{3}, become polynomial-time solvable once any planar graph is excluded as a minor.

The price for this rich theory is sparsity.
Every proper minor-closed class excludes some complete graph as a minor, and therefore has bounded average degree;
in particular, it cannot contain arbitrarily large cliques.
To say anything about dense graphs, one has to work with a containment relation that does not allow deleting edges.
The most natural candidate is the induced subgraph relation.
However, \emph{hereditary classes}, i.e., classes closed under taking induced subgraphs, seem to be too diverse to share strong common properties. (A notable exception is the Erd\H{o}s--Hajnal conjecture~\cite{ErdosHajnal1989}.)

\emph{Induced minors} offer a middle ground between these two worlds.
A graph $H$ is an induced minor of a graph $G$ if $H$ can be obtained from $G$ by deleting vertices and contracting edges; in contrast to minors, deleting edges is not allowed.
Consequently, classes excluding a fixed induced minor may be dense.
For instance, if $H$ is not a complete graph, then every complete graph is $H$-induced-minor-free.
At the same time, such classes are much more restricted than general hereditary classes, and some of the uniform behaviour of minor-free classes appears to survive.
For example, Korhonen and Lokshtanov~\cite{KorhonenLokshtanov2024} proved that, for every fixed graph $H$, every $m$-edge $H$-induced-minor-free graph has a balanced separator of size $\Oh_H(\sqrt m)$, which is a direct analogue of the separator theorem of Alon, Seymour, and Thomas~\cite{AlonSeymourThomas1990}.
More recently, Bonnet and Chang~\cite{BonnetChangQPTAS2026} obtained a quasipolynomial-time approximation scheme (QPTAS) for \MWIS in every class excluding a fixed induced minor.
Their algorithm relies on balanced separators of size sublinear in the number of edges, such as those provided by Korhonen and Lokshtanov, and can be seen as an induced-minor analogue of the approximation scheme for minor-free graphs.

\paragraph{Excluding a planar induced minor.}
Given the role of planar graphs in the theory of minors, it is natural to ask whether they also mark the boundary of tractability for induced minors.
Dallard, Milani\v{c}, and \v{S}torgel~\cite{dallard2022secondpaper} asked whether \MWIS is polynomial-time solvable in $H$-induced-minor-free graphs for every planar graph $H$.
Independently, Gartland and Lokshtanov proposed a broader set of conjectures along the same lines; see Gartland's thesis~\cite{GartlandThesis}.
We focus on the following prediction, which covers both independent sets and colorings.

\begin{conjecture}\label{conj:algorithmic}
    For every planar graph $H$, \MWIS and \lcol{3} are polynomial-time solvable in $H$-induced-minor-free graphs.
\end{conjecture}

The planarity assumption is necessary: if $H$ is not planar, then every planar graph is $H$-induced-minor-free, while both \MIS and \col{3} are \NP-hard on planar graphs~\cite{GAREY1976237}.
On the other hand, \cref{conj:algorithmic} would generalize a long line of algorithmic results.
Indeed, a graph contains the path $P_t$ as an induced minor if and only if it contains it as an induced subgraph, and it contains the cycle $C_t$ as an induced minor if and only if it has an induced cycle of length at least $t$.
Thus the polynomial-time algorithms for \MWIS in $P_5$-free and $P_6$-free graphs~\cite{DBLP:conf/soda/LokshantovVV14,grzesik2020polynomialtime}, as well as the algorithms for several further planar graphs $H$~\cite{dallard2022secondpaper,DBLP:journals/algorithmica/BonnetDGTW26}, are special cases of \cref{conj:algorithmic}.
For general paths and cycles, only quasipolynomial-time algorithms are known~\cite{DBLP:conf/focs/GartlandL20,DBLP:conf/sosa/PilipczukPR21,DBLP:conf/stoc/GartlandLPPR21}.
Hence, \cref{conj:algorithmic} would be interesting, and highly nontrivial, even with ``polynomial'' relaxed to ``quasipolynomial''.

Why should \cref{conj:algorithmic} be true?
Gartland and Lokshtanov~\cite{GartlandThesis} proposed a structural explanation in the spirit of the Grid Minor Theorem.
Since classes excluding an induced minor may contain arbitrarily large cliques, one cannot hope for balanced separators of bounded size.
Instead, one can measure a separator by the number of vertices needed to dominate it.
For a function $f$, we say that a class $\cC$ of graphs admits \emph{$f(n)$-dominated balanced separators} if for every $n$-vertex graph $G \in \cC$ and every weight function $w\colon V(G)\to\mathbb{R}_{\geq 0}$, there exists a set $X$ of at most $f(n)$ vertices such that $N[X]$ is a balanced separator of $(G,w)$.
If $f$ is a constant~$k$, we speak of \emph{$k$-dominated balanced separators}, and we write \emph{$\Oh(1)$-dominated} if such a constant exists.
Such separators may be large, but they are structurally simple: they are covered by few closed neighborhoods.
They are also the engine behind the quasipolynomial-time algorithms for $P_t$-free graphs: by the Gy\'arf\'as path argument, the class of $P_t$-free graphs admits $(t-1)$-dominated balanced separators~\cite{DBLP:conf/focs/GartlandL20,DBLP:conf/sosa/PilipczukPR21}.
Gartland and Lokshtanov believed that dominated balanced separators are the key to the tractability predicted by \cref{conj:algorithmic}, and conjectured the following.

\begin{conjecture}[Gartland--Lokshtanov]\label{conj:dominated-separators}
    For every planar graph $H$, there exists a constant $k_H$ such that the class of $H$-induced-minor-free graphs admits $k_H$-dominated balanced separators.
\end{conjecture}

Again, planarity is necessary: grids have bounded degree, but their balanced separators have size $\Omega(\sqrt n)$, so they cannot be dominated by few vertices.
\Cref{conj:dominated-separators} has been confirmed for several families of planar graphs $H$~\cite{dallard2022secondpaper,DBLP:journals/algorithmica/BacsoLMPTL19,DBLP:journals/siamcomp/ChudnovskyPPT24,DBLP:journals/corr/abs-2512-12329,CzyzewskaMPRR2026}.

\paragraph{The refutation.}
Recently, Bonnet and Chang~\cite{BonnetChang2026} refuted the independent-set part of \cref{conj:algorithmic}: \MIS remains \NP-hard in graphs excluding the $5\times5$ grid as an induced minor.
As their reduction runs in polynomial time, it also rules out quasipolynomial-time algorithms, unless $\NP\subseteq\QP$, where \QP denotes the class of problems solvable in quasipolynomial time.
Thus, excluding a planar induced minor does not make \MIS tractable. 
Moreover, the relaxed, quasipolynomial version of \cref{conj:algorithmic} fails as well.

This leaves the research program in an unclear state, and raises several natural questions.
For which planar graphs $H$ does the hardness hold, and how far are they from the graphs $H$ for which efficient algorithms are known?
Is the construction of Bonnet and Chang essentially the only obstruction?
Does the coloring part of \cref{conj:algorithmic} survive?
And, perhaps most importantly, what does the failure of \cref{conj:algorithmic} tell us about \cref{conj:dominated-separators}, which was meant to explain it?
In this paper, we address all these questions.

\subsection*{Our results}

\begin{figure}[htbp]
    \centering
    % The four minimal trees with an asteroidal triple of nonpaths.
% Panel k has k claws attached at their centres and 3-k at leaves.
\begin{tikzpicture}[
    vertex/.style={circle,draw,fill=white,inner sep=0pt,minimum size=4pt,line width=.65pt},
    edge/.style={line width=.65pt},
    every node/.style={font=\small}
]
    \foreach \k in {0,1,2,3} {
        \pgfmathsetmacro{\panelx}{3.9*\k}
        \begin{scope}[xshift=\panelx cm]
            \begin{scope}[scale=.75]
                \foreach \arm/\angle in {
                    1/0,
                    2/120,
                    3/240} {
                    \begin{scope}[rotate=\angle]
                        \draw[edge] (0,0) -- (0,.64) -- (0,1.25);
                        \node[vertex] at (0,.64) {};
                        \ifnum\arm>\k
                            % The connecting path meets a leaf of the claw.
                            \draw[edge,draw=orange!85!black,line width=0.5mm]
                                (0,1.25) -- (0,1.85) -- (-.45,2.2)
                                (0,1.85) -- (.45,2.2);
                            \foreach \point in {(0,1.25),(0,1.85),(-.45,2.2),(.45,2.2)}
                                \node[vertex] at \point {};
                        \else
                            % The connecting path meets the centre of the claw.
                            \draw[edge,draw=orange!85!black,line width=0.5mm]
                                (0,1.25) -- (-.52,1.64)
                                (0,1.25) -- (0,2.1)
                                (0,1.25) -- (.52,1.64);
                            \foreach \point in {(0,1.25),(-.52,1.64),(0,2.1),(.52,1.64)}
                                \node[vertex] at \point {};
                        \fi
                    \end{scope}
                }
                \node[vertex] at (0,0) {};
            \end{scope}
            \node at (0,-1.5) {$F_{\k}$};
        \end{scope}
    }
\end{tikzpicture}
    \caption{The four $16$-vertex trees that are not induced minors of any loom grid or triangle-free loom grid.
    In each tree, three claws (highlighted) are attached to the ends of a subdivided claw; in $F_k$, exactly $k$ claws are attached at their centres.}
    \label{fig:grid-trees}
\end{figure}
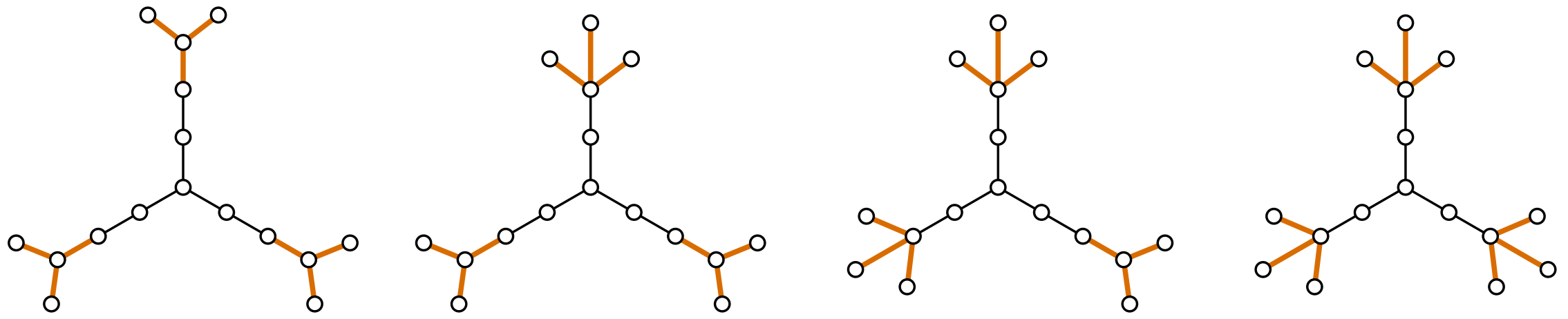

\paragraph{What loom grids exclude.}
We call the host graphs constructed by Bonnet and Chang \emph{loom grids}, and denote their family by $\cB$; see \cref{sec:grids} for the definition.
Bonnet and Chang showed that no induced minor of a loom grid contains three pairwise antiadjacent cycles such that any two of them are joined by a path avoiding the closed neighborhood of the third.
We observe that the cycles in this statement can be replaced by arbitrary connected graphs that are not paths.
As a consequence, loom grids exclude planar graphs that are much simpler than the $5\times5$ grid.
In particular, none of the four $16$-vertex trees in \cref{fig:grid-trees} is an induced minor of a loom grid.
We also show that loom grids exclude the graph $\subKfour$ obtained by subdividing every edge of $K_4$ exactly once, which is not covered by the above argument.

\paragraph{Maximum Independent Set.}
Combining the hardness result of Bonnet and Chang with the above observations, we obtain that for each tree $H$ in \cref{fig:grid-trees}, \MIS is \NP-hard in $H$-induced-minor-free graphs. 
This should be contrasted with the quasipolynomial-time algorithms for the cases when $H$ is a path or a subdivided claw~\cite{DBLP:conf/focs/GartlandL20,DBLP:conf/sosa/PilipczukPR21,DBLP:conf/stoc/GartlandLMPPR24}: each tree in \cref{fig:grid-trees} has only four vertices of degree larger than two.

One might hope that the construction of Bonnet and Chang is essentially the only obstruction, so that \cref{conj:algorithmic} holds at least for those planar graphs $H$ that are induced minors of all sufficiently large loom grids.
We show that this is not the case.
We introduce a second family $\cT$ of \emph{triangle-free loom grids}, and show that it provides a further, independent obstruction: \MIS is \NP-hard in the hereditary closure of $\cT$.
Our reduction is a slight modification of the one of Bonnet and Chang, and the same construction works for $\cB$ as well: the hard instance for $\cT$ is obtained from the hard instance for $\cB$ by deleting some edges.
Hence, we obtain the following two theorems; the first one is due to Bonnet and Chang~\cite{BonnetChang2026}, and our proof of it is self-contained, while the second one is new.

\begin{restatable}{theorem}{thmmwisB}
    \MIS is \NP-hard in the hereditary closure of $\cB$.
\end{restatable}

\begin{restatable}{theorem}{thmmwisT}
    \MIS is \NP-hard in the hereditary closure of $\cT$.
\end{restatable}

The two families differ in the planar graphs they exclude.
We exhibit a $16$-vertex planar graph $H_1$ that is an induced subgraph of every sufficiently large loom grid, but not an induced minor of any triangle-free loom grid.
Thus, \MIS remains \NP-hard in $H_1$-induced-minor-free graphs, even though this class excludes all large loom grids.

\paragraph{3-Coloring.}
The result of Bonnet and Chang leaves open whether the coloring part of \cref{conj:algorithmic} might survive.
We show that it does not, even for ordinary \col{3}.

\begin{restatable}{theorem}{thmcolB}
    \col{$3$} is \NP-hard in the hereditary closure of $\cB$.
\end{restatable}

\begin{restatable}{theorem}{thmcolT}
    \col{$3$} is \NP-hard in the induced minor closure of $\cT$.
\end{restatable}

For $\cT$, passing to the induced minor closure is essential: every triangle-free loom grid, and hence every graph in its hereditary closure, is $3$-colorable.
Consequently, \col{3}, and thus also \lcol{3}, is \NP-hard in $H$-induced-minor-free graphs whenever $H$ is one of the trees in \cref{fig:grid-trees}, the graph $\subKfour$, or the graph $H_1$.
Again, this rules out quasipolynomial-time algorithms unless $\NP\subseteq\QP$.
Together with the result of Bonnet and Chang, this refutes \cref{conj:algorithmic} in full, even in its quasipolynomial version.

\paragraph{Dominated balanced separators.}
Since \cref{conj:dominated-separators} was proposed as the structural reason behind \cref{conj:algorithmic}, one might expect it to fail as well.
We show that the two conjectures in fact come apart.

A \emph{heeled caterpillar} is a graph obtained from a caterpillar by adding two pendant vertices to every leaf, and then subdividing each edge arbitrarily many times.
As our next result, we confirm \cref{conj:dominated-separators} for all heeled caterpillars.

\begin{restatable}{theorem}{thmdbscaterpillar}\label{thm:dbscaterpillar}
    For every heeled caterpillar $H$, the class of $H$-induced-minor-free graphs admits $\Oh(1)$-dominated balanced separators.
\end{restatable}

The constant depends on $H$.
The tree $F_0$ in \cref{fig:grid-trees} is a heeled caterpillar: it is obtained from a claw by adding two pendant vertices to each leaf and subdividing each edge of the claw twice.
Hence, the class of $F_0$-induced-minor-free graphs admits $\Oh(1)$-dominated balanced separators, yet \MIS is \NP-hard in this class.
In other words, \cref{conj:dominated-separators} holds for this $H$, while \cref{conj:algorithmic} fails.
In particular, unless $\NP\subseteq\QP$, the existence of $\Oh(1)$-dominated balanced separators in a hereditary class does not by itself yield quasipolynomial-time algorithms for \MIS.

\paragraph{Organization.}
After introducing notation in \cref{sec:prelims}, we define loom grids and triangle-free loom grids and analyze their excluded induced minors in \cref{sec:grids}.
The hardness results for \MIS and \col{3} are proved in \cref{sec:NPmis} and \cref{sec:coloring}, respectively.
\Cref{thm:dbscaterpillar} is proved in \cref{sec:dbs}.
We conclude with open problems in \cref{sec:conclusion}.

\section{Preliminaries}\label{sec:prelims}
All graphs considered in this paper are simple, finite and undirected. A vertex set of a graph $G$ is denoted by $V(G)$ and the set of edges by $E(G)$. An edge connecting vertices $u$ and $v$ is denoted by $uv$. 
A set of vertices adjacent to a vertex $v$ is denoted by $N(v)$ (so called \emph{open neighborhood}). By $N[v]$ we denote the set $N(v)\cup \{v\}$. We also set $N(A)=\bigcup_{v\in A} N(v) \setminus A$ and $N[A]=\bigcup_{v\in A} N[v]$.
For two sets $A,B \subseteq V(G)$, we say that $A$ is \emph{anticomplete} to $B$ if no edge of $G$ has one endpoint in $A$ and the other endpoint in $B$.
Note that $A$ is disjoint from and anticomplete to $B$ if and only if $A \cap N[B] = \emptyset$.
We say that set $A\subseteq V(G)$ \emph{dominates} $B\subseteq V(G)$ if $B\subseteq N[A]$.
We denote a complete graph on $n$ vertices by $K_n$ and a complete bipartite graph with sides of bipartition of size $t$ by $K_{t,t}$. 
A \emph{claw} is a tree on four vertices, with one vertex of degree $3$. 
A path is denoted as a sequence of vertices $v_1v_2\dots v_n$. The \emph{length} of the path is the number of edges in it.

Given graphs $G$ and $H$, we say that \emph{induced minor model} $\mathcal{M}$ of $H$ in $G$ is a function assigning each vertex $v$ of $H$ a connected vertex set $X_v$ of $G$ such that:
\begin{enumerate}
    \item for every  distinct $u,v\in V(G)$ sets $X_u$ and $X_v$ are disjoint;
    \item for each edge $uv\in E(H)$  sets $X_u$ and $X_v$ are adjacent;
    \item for every non-edge $uv\notin E(H)$ sets $X_u$ and $X_v$ are antiadjacent.
\end{enumerate}
We say that $H$ is an \emph{induced minor} of $G$ if there exists an induced minor model $\mathcal{M}$ of $H$ in $G$. 
For a model $\mathcal{M}$, let $V(\mathcal{M})$ denote the union of its branch sets.

\section{Loom grids}\label{sec:grids}
In this section we recall a family $\cB$ of graphs considered by Bonnet and Chang~\cite{BonnetChang2026},
which we call \emph{loom grids}, and introduce the family $\cT$ of \emph{triangle-free loom grids}.
We then analyze some (families of) planar graphs excluded as induced minors by these families.

\subsection{Definitions of loom grids.}
Let $r,c$ be positive integers.
The \emph{loom grid} $B_{r,c}$ is defined as follows:
\begin{align*}
    V(B_{r,c})  = & \bigl\{ (i,j,s) \mid i \in [r], j \in [c], s \in \{-,+\} \bigr\},\\
    E(B_{r,c})  = & \bigl\{ (i,j,-)(i',j,+) \mid i,i' \in [r], j \in [c] \bigr\} \\
    & \cup  \bigl\{ (i,j,s)(i,j+1,s') \mid i \in [r], j \in [c-1], s,s' \in \{-,+\} \bigr\}.
\end{align*}
We call $i$ the row, $j$ the column, and $s$ the side.
The vertices with fixed row $i$ and column $j$ form the \emph{cell} $(i,j)$.
Edges within a cell are called \emph{internal}.
Edges joining different cells of the same column are called \emph{vertical}, and edges between consecutive columns are called \emph{horizontal}.
Whenever we talk about vertices/sets being to the left or right from some other vertex or sets, we refer to the column index of the vertices (with smaller indices being to the left).
The \emph{triangle-free loom grid} $T_{r,c}$ is defined as follows:
\begin{align*}
    V(T_{r,c}) & = \bigl\{ (i,j,-) \mid i \in [r], j \in [c] \bigr\}
        \cup \bigl\{ (i,j,+) \mid i \in [r], \text{ odd } j \in [c] \bigr\},\\
    E(T_{r,c}) & = \bigl\{ (i,j,-)(i',j,+) \mid i,i' \in [r], i \ne i', \text{ odd } j \in [c] \bigr\} \\
        & \cup \bigl\{ (i,j,s)(i,j+1,s') \mid (i,j,s),(i,j+1,s') \in V(T_{r,c}) \bigr\}.
\end{align*}

Equivalently, $T_{r,c}$ is obtained from $B_{r,c}$ by deleting all vertices $(i,j,+)$ with even $j$ and all remaining internal edges.
In particular, each odd column induces $K_{r,r}$ with a perfect matching removed.
The graph $T_{r,c}$ is triangle-free: a triangle cannot lie within a column, and a triangle using two columns would need an internal edge in an odd column.
Define
\[
\cB = \{ B_{r,c} \mid r,c \geq 1 \}, \qquad
\cT = \{ T_{r,c} \mid r,c \geq 1 \}.
\]

\begin{figure}[htbp]
    \centering
    \begin{tikzpicture}[
    x=1cm,y=1cm,
    vertex/.style={circle,draw=black,fill=white,inner sep=0pt,minimum size=4pt,line width=.65pt},
    vertical edge/.style={draw=blue!65!black,line width=.45pt,opacity=.7},
    horizontal edge/.style={draw=black,line width=.55pt},
    internal edge/.style={draw=red!75!black,line width=.65pt},
    every node/.style={font=\small,inner sep=2pt},
]
    % kind=0 gives B_{5,5}; kind=1 deletes the required vertices and edges for T_{5,5}.
    \foreach \familyletter/\kind/\panelx in {B/0/0,T/1/8.1} {
        \begin{scope}[xshift=\panelx cm]
            \node at (2.8,1.25) {$\familyletter_{5,5}$};
            \foreach \column in {1,...,5} {
                \pgfmathtruncatemacro{\hasplus}{(\kind==0)||(mod(\column,2)==1)}
                \node at ({1.4*(\column-1)},.8) {$\column$};
                \node[font=\scriptsize] at ({1.4*(\column-1)-.18},.43) {$-$};
                \ifnum\hasplus=1\relax
                    \node[font=\scriptsize] at ({1.4*(\column-1)+.18},.43) {$+$};
                \fi
                \foreach \row in {1,...,5} {
                    \coordinate (\familyletter-\row-\column-minus) at
                        ({1.4*(\column-1)-.18},{-.85*(\row-1)+.10});
                    \ifnum\hasplus=1\relax
                        \coordinate (\familyletter-\row-\column-plus) at
                            ({1.4*(\column-1)+.18},{-.85*(\row-1)-.10});
                    \fi
                }
            }

            % Vertical edges join opposite sides in distinct rows of one column.
            \foreach \column in {1,...,5} {
                \pgfmathtruncatemacro{\hasplus}{(\kind==0)||(mod(\column,2)==1)}
                \ifnum\hasplus=1\relax
                    \foreach \row in {1,...,5} {
                        \foreach \otherrow in {1,...,5} {
                            \ifnum\row=\otherrow\relax\else
                                \draw[vertical edge] (\familyletter-\row-\column-minus)
                                    -- (\familyletter-\otherrow-\column-plus);
                            \fi
                        }
                    }
                \fi
            }

            % Join all surviving vertices in consecutive cells of each row.
            \foreach \column in {1,...,4} {
                \pgfmathtruncatemacro{\nextcolumn}{\column+1}
                \pgfmathtruncatemacro{\sourceplus}{(\kind==0)||(mod(\column,2)==1)}
                \pgfmathtruncatemacro{\targetplus}{(\kind==0)||(mod(\nextcolumn,2)==1)}
                \foreach \row in {1,...,5} {
                    \draw[horizontal edge] (\familyletter-\row-\column-minus)
                        -- (\familyletter-\row-\nextcolumn-minus);
                    \ifnum\sourceplus=1\relax
                        \draw[horizontal edge] (\familyletter-\row-\column-plus)
                            -- (\familyletter-\row-\nextcolumn-minus);
                    \fi
                    \ifnum\targetplus=1\relax
                        \draw[horizontal edge] (\familyletter-\row-\column-minus)
                            -- (\familyletter-\row-\nextcolumn-plus);
                        \ifnum\sourceplus=1\relax
                            \draw[horizontal edge] (\familyletter-\row-\column-plus)
                                -- (\familyletter-\row-\nextcolumn-plus);
                        \fi
                    \fi
                }
            }

            % Internal edges are present only in B; draw them above the other edges.
            \ifnum\kind=0\relax
                \foreach \column in {1,...,5} {
                    \foreach \row in {1,...,5} {
                        \draw[internal edge] (\familyletter-\row-\column-minus)
                            -- (\familyletter-\row-\column-plus);
                    }
                }
            \fi

            % Draw vertices last so that edges do not pass through their interiors.
            \foreach \row in {1,...,5}
                \node at (-.7,{-.85*(\row-1)}) {$\row$};
            \foreach \column in {1,...,5} {
                \pgfmathtruncatemacro{\hasplus}{(\kind==0)||(mod(\column,2)==1)}
                \foreach \row in {1,...,5} {
                    \node[vertex] at (\familyletter-\row-\column-minus) {};
                    \ifnum\hasplus=1\relax
                        \node[vertex] at (\familyletter-\row-\column-plus) {};
                    \fi
                }
            }
        \end{scope}
    }
\end{tikzpicture}
    \caption{A loom grid $B_{5,5}$ (left) and a triangle-free loom grid $T_{5,5}$ (right), with rows and columns numbered and the two sides marked $-$ and $+$.
    Blue edges are vertical, black edges are horizontal, and red edges are internal.}
    \label{fig:grid-families}
\end{figure}
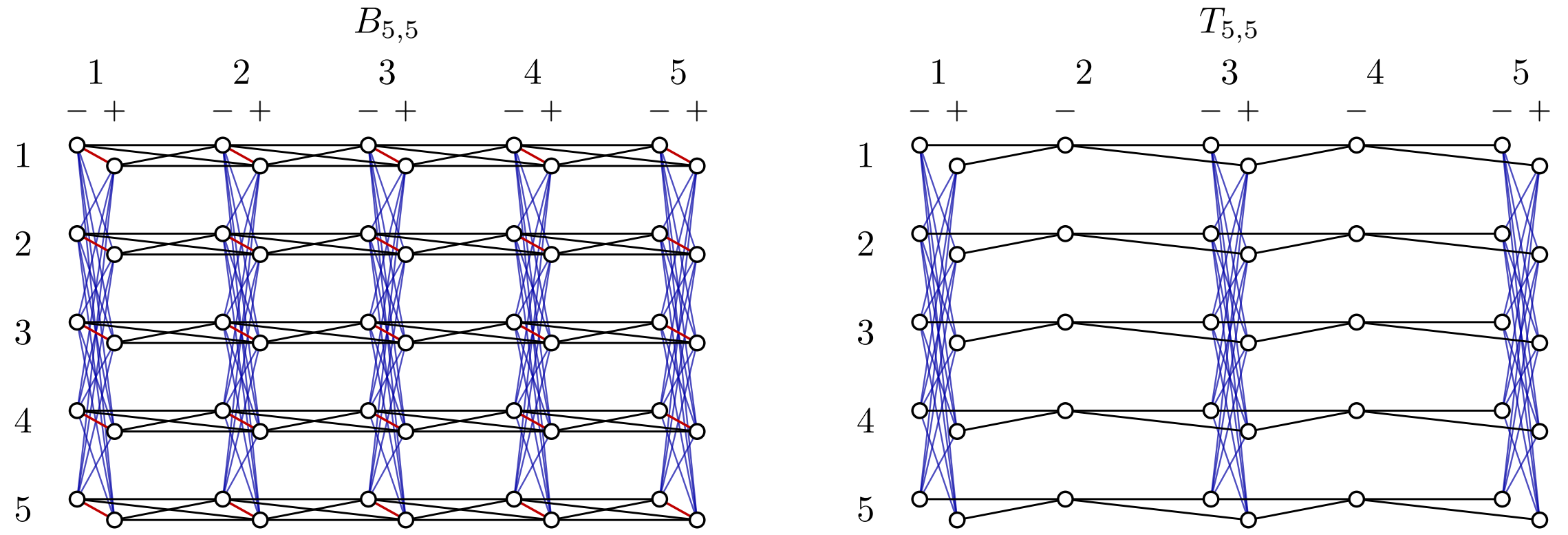

The graphs $B_{5,5}$ and $T_{5,5}$ are shown in \cref{fig:grid-families}.

\subsection{Induced minors of loom grids.}

Bonnet and Chang~\cite{BonnetChang2026} proved that no induced minor of a graph from $\cB$ has three pairwise antiadjacent cycles, each two of them joined by a path avoiding the closed neighborhood of the third one.
We extend the idea of Bonnet and Chang by replacing cycles with arbitrary connected graphs that are not paths.

An \emph{asteroidal triple of nonpaths} in a graph $G$ consists of three pairwise disjoint and antiadjacent vertex sets, each inducing a connected graph that is not a path, such that any two sets are joined by a path avoiding the closed neighborhood of the third set.

We first record some properties of the columns that will be useful later.
A vertex set is \emph{mixed} in a column if it contains vertices on both sides of that column.
Note that for graphs in $\cT$, a set may be mixed only on odd columns, as even columns have only one side.

\begin{lemma}\label{lem:grid-barrier}
    Let $G \in \cB \cup \cT$. The following hold.
    \begin{enumerate}[(a),ref=(\alph*)]
        \item\label{item:grid-barrier-path} Every nonempty connected vertex set with no mixed column induces a path contained in one row.
        \item\label{item:grid-barrier-components} If $P\subseteq V(G)$ consists of two vertices on opposite sides of a column $j$, then every component of $G-N[P]$ meeting column $j$ has at most two vertices.
        \item\label{item:grid-barrier-nonpath} With $P$ and $j$ as in \ref{item:grid-barrier-components}, every connected set with at least three vertices that is antiadjacent to $P$ lies strictly on one side of column $j$.
    \end{enumerate}
\end{lemma}
\begin{proof}
    For \ref{item:grid-barrier-path}, let $U$ be a nonempty connected vertex set with no mixed column.
    Then $G[U]$ has neither vertical nor internal edges.
    Thus all its vertices lie in one row, with at most one vertex in each column, and its columns are consecutive.
    It therefore induces a path.

    For \ref{item:grid-barrier-components}, let $P$ consist of two vertices on opposite sides of column $j$.
    In $B_{r,c}$, an opposite-side pair dominates its entire column so the claim holds vacuously.
    In $T_{r,c}$, column $j$ is necessarily odd. If $P$ lies in the same row, the argument is the same as for $B_{r,c}$. So suppose that the vertices of $P$ are $(i,j,-)$ and $(k,j,+)$, where $i \ne k$.
    The only vertices of column $j$ outside its closed neighborhood are $(k,j,-)$ and $(i,j,+)$.
    These form an edge, and all their horizontal neighbors are deleted.
    Hence they form a two-vertex component of $G-N[P]$.

    For \ref{item:grid-barrier-nonpath}, by \ref{item:grid-barrier-components}, a connected set with at least three vertices antiadjacent to $P$ avoids column $j$, so connectivity places it strictly on one side of that column.
\end{proof}

\begin{lemma}\label{lem:grid-asteroidal}
    No induced minor of a graph from $\cB \cup \cT$ contains an asteroidal triple of nonpaths.
\end{lemma}
\begin{proof}
    Suppose that $(X_v \mid v \in V(H))$ is an induced minor model of a graph $H$ containing such a triple $A_1,A_2,A_3$ in a host $G \in \cB \cup \cT$.
    For $i \in \{1,2,3\}$, put $U_i=\bigcup_{v \in A_i}X_v$.
    The sets $U_1,U_2,U_3$ are connected and pairwise antiadjacent.
    Furthermore, $G[U_i]$ is not a path, since $H[A_i]$ is its induced minor and every induced minor of a path is a disjoint union of paths.
    By \cref{lem:grid-barrier}\ref{item:grid-barrier-path}, each $U_i$ is mixed in column $j_i$.
    By \cref{lem:grid-barrier}\ref{item:grid-barrier-nonpath}, these columns are distinct.
    By symmetry, assume that $j_1<j_2<j_3$.
    Again by \cref{lem:grid-barrier}\ref{item:grid-barrier-nonpath}, $U_1$ lies strictly to the left of $j_2$ and $U_3$ strictly to its right.

    Let $Q$ be an $A_1$--$A_3$ path in $H-N[A_2]$.
    Its image $U=\bigcup_{v \in V(Q)}X_v$ is connected, has at least three vertices, and is antiadjacent to $U_2$, yet meets columns on both sides of $j_2$.
    This contradicts \cref{lem:grid-barrier}\ref{item:grid-barrier-nonpath} applied to an opposite-side pair in $U_2$ in column $j_2$.
\end{proof}

In particular, \cref{lem:grid-asteroidal} excludes some trees as induced minors of graphs from $\cB\cup\cT$.
The four trees in \cref{fig:grid-trees} are precisely the trees minimal under induced minors among those containing an asteroidal triple of nonpaths.
Indeed, given such a triple in a tree, choose an induced claw in each nonpath, retain the claws and their connecting subtree, and shorten its three arms until the central branch vertex has distance two from each claw.
Each arm meets its claw either at its centre or at a leaf, giving the four possibilities shown.
Each resulting tree has $16$ vertices and is minimal: any asteroidal triple of nonpaths in a tree requires at least twelve vertices in the three nonpaths, together with the central branch vertex and one further vertex on each of its three arms.

\cref{lem:grid-asteroidal} does not account for every planar graph excluded by $\cB$.
Let $\subKfour$ be the graph obtained from $K_4$ by subdividing every edge exactly once.
Call the four original vertices $a,b,c,d$, and denote the subdivision vertex on the edge $uv$ by $z_{uv}=z_{vu}$.
Thus $\subKfour$ has ten vertices.

\begin{lemma}\label{lem:grid-H0}
    The graph $\subKfour$ is planar and contains no asteroidal triple of nonpaths, but no graph from $\cB$ contains $\subKfour$ as an induced minor.
\end{lemma}
\begin{proof}    
    Planarity follows from planarity of $K_4$, and triangle-freeness implies that every connected induced nonpath has at least four vertices.
    Thus the ten vertices of $\subKfour$ cannot even contain three pairwise disjoint nonpaths.

    Suppose that $(X_v \mid v \in V(\subKfour))$ is an induced minor model in $B_{r,q}$, and put $W=\bigcup_{v \in V(\subKfour)}X_v$.
    In this proof, a column is \emph{active} if $W$ is mixed in that column, and \emph{inactive} otherwise.

    \begin{claim}\label{clm:grid-subkfour-original}     
        For each $v \in \{a,b,c,d\}$, the set $X_v$  meets at least one active column.
    \end{claim}
    \begin{claimproof}
    Suppose that $X_v$ avoids all active columns. By \cref{lem:grid-barrier}\ref{item:grid-barrier-path},
    $X_v$ induces a~path in one row, say $i$.
    Let $X_v$ meet columns $\ell,\ell+1,\ldots,t$.
    These columns are inactive, so all vertices of $W$ in each of them lie on the same side as the vertex of $X_v$.
    Hence no vertical or internal edge joins $X_v$ to another branch set, and no other branch set meets a~cell occupied by $X_v$.
    
    Since horizontal edges join consecutive cells in the same row, every neighbor of $X_v$ belonging to another branch set must lie in cell $(i,\ell-1)$ or $(i,t+1)$, provided that these cells exist.
    Suppose that two distinct branch sets $X_u,X_w$ contain neighbors of $X_v$ in the same one of these two cells.
    Each cell induces an internal edge, so $X_u$ and $X_w$ are also adjacent.
    Thus $u,v,w$ form a triangle in $\subKfour$, a~contradiction.
    Consequently, $v$ has at most two neighbors in $\subKfour$. This is a contradiction, as $a,b,c,d$ are of degree 3 in $\subKfour$.
    \end{claimproof}

    For each active column $j$, define
    \[
        S_j=\{v \in V(\subKfour) \mid X_v \text{ meets column }j\}.
    \]

    \begin{claim}\label{clm:grid-subkfour-star}
        For every active column $j$, $\subKfour[S_j]$ is a star, and all branch sets $X_v$ for $v \notin S_j$ lie strictly on the same side of column $j$.
    \end{claim}
    \begin{claimproof}
        If some branch set $X_v$ meets both sides of column $j$, then $v$ is adjacent to every vertex of $S_j\setminus\{v\}$, so triangle-freeness implies that $\subKfour[S_j]$ is a star.
        Otherwise, partition $S_j$ according to the side of column $j$ met by the corresponding branch set.
        The two parts are nonempty and complete to each other.
        Triangle-freeness makes each part independent, and the absence of $4$-cycles forces one part to be a singleton.
        Thus $\subKfour[S_j]$ is again a star.

        It is straightforward to verify that removing a star from $\subKfour$ leaves a connected graph.
        Thus the union of all branch sets $X_v$ with $v\notin S_j$ is connected as well.
        By the definition of $S_j$, this union avoids column $j$.
        Since edges of $B_{r,q}$ join vertices in the same or consecutive columns, a connected set avoiding column $j$ lies strictly on one side of it.
    \end{claimproof}

    \begin{claim}\label{clm:grid-subkfour-columns}
        There are active columns $j<k$, with no active column strictly between them, and vertices $a',b',c',d'$ with $\{a',b',c',d'\}=\{a,b,c,d\}$ such that
        \[
            S_j=\{a',z_{a'b'},b'\},\qquad S_k=\{c',z_{c'd'},d'\},
        \]
        and every branch set $X_v$ with $v\notin S_j\cup S_k$ lies strictly between $j$ and $k$.
    \end{claim}
    \begin{claimproof}
        By \cref{clm:grid-subkfour-star}, all branch sets avoiding a given active column lie strictly on the same side of it.
        This side contains another active column, since a star in $\subKfour$ contains at most two among $a,b,c,d$,
        and each of their branch sets meets an active column.
        The branch sets outside the first active column therefore lie to its right, and those outside the last lie to its left.
        Let $k$ be the first active column such that all branch sets avoiding it lie strictly to its left, and let $j$ be the preceding active column.
        Then all branch sets avoiding $j$ lie strictly to its right, and no active column lies strictly between $j$ and $k$.
        Every branch set $X_v$ with $v\notin S_j\cup S_k$ lies strictly between $j$ and $k$, so meets no active column.
        Thus, by \cref{clm:grid-subkfour-original}, $S_j\cup S_k$ contains all four original vertices.
        A star contains at most two of them, with equality only if it is of the form $\{u,z_{uv},v\}$ for $u,v \in \{a,b,c,d\}$.
        Denote the two original vertices in $S_j$ by $a',b'$ and those in $S_k$ by $c',d'$; these are four distinct vertices and give the required forms of $S_j$ and $S_k$.
    \end{claimproof}

    Fix columns $j<k$ and vertices $a',b',c',d'$ as in \cref{clm:grid-subkfour-columns}.
    By symmetry, assume that $a'=a$, $b'=b$, $c'=c$, and $d'=d$.
    The antiadjacent sets $X_a,X_b$ both meet column $j$ and lie strictly left of column $k$.
    At most one of $X_a,X_b$ can be mixed in a column before $j$: at the rightmost of two such columns, an opposite-side pair in one set would dominate a vertex of the other on its way to $j$.
    Neither set is mixed in column $j$, and there are no active columns between $j$ and $k$.
    Thus one, say $X_a$, is a path in one row by \cref{lem:grid-barrier}\ref{item:grid-barrier-path}.
    The branch sets $X_{z_{ac}},X_{z_{ad}}$ lie strictly between $j$ and $k$.
    Since these columns are inactive, no vertical or internal edge joins either set to $X_a$, and neither set can meet a cell occupied by $X_a$.
    Any adjacency to $X_a$ must therefore occur in a cell immediately beyond one of its endpoints.
    Both sets lie strictly to the right of column $j$, which $X_a$ meets, so they must both meet the cell immediately beyond its right endpoint.
    This contradicts their antiadjacency, since each cell of $B_{r,q}$ induces an internal edge.
\end{proof}

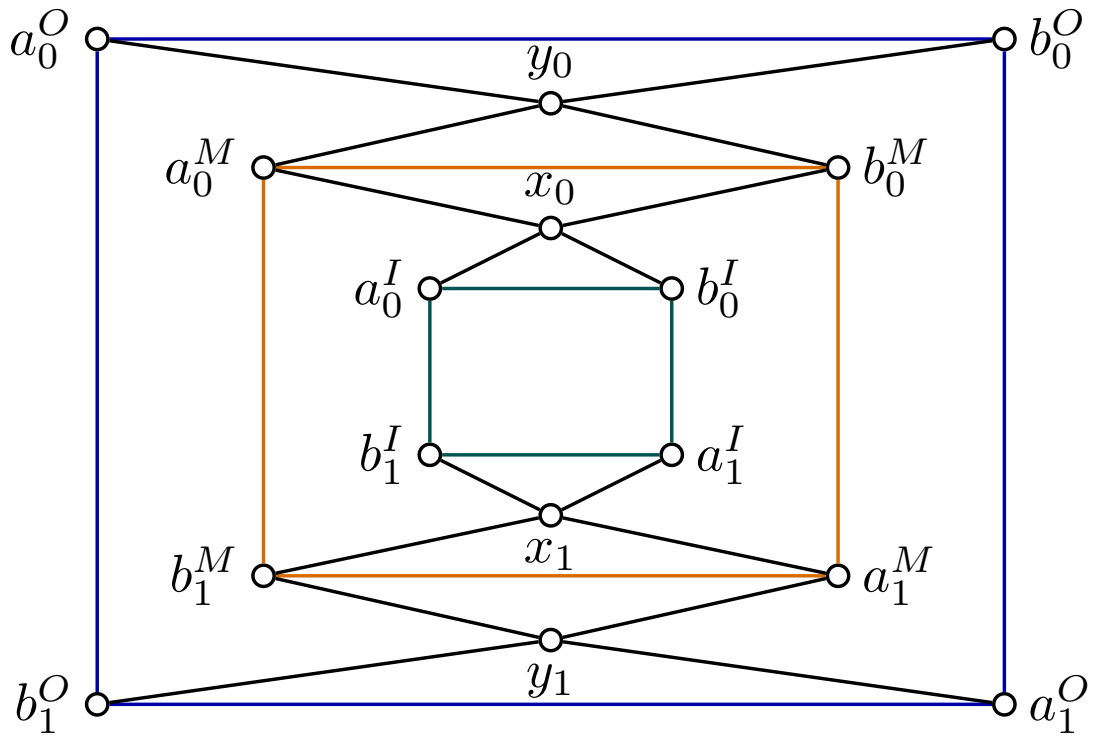
\begin{figure}[htbp]
    \centering
    \begin{tikzpicture}[
    x=1cm,y=1cm,
    vertex/.style={circle,draw=black,fill=white,inner sep=0pt,minimum size=4pt,line width=.65pt},
    edge/.style={line width=.65pt},
    every label/.style={font=\small,inner sep=2pt},
]
    % Each cycle is a_0^C b_0^C a_1^C b_1^C a_0^C.
    \foreach \cycle/\width/\height/\cyclecolour in {
        I/.8/.55/teal!65!black,
        M/1.9/1.35/orange!85!black,
        O/3/2.2/blue!65!black} {
        \node[vertex,label=left:{$a_0^{\cycle}$}] (a0\cycle) at (-\width,\height) {};
        \node[vertex,label=right:{$b_0^{\cycle}$}] (b0\cycle) at (\width,\height) {};
        \node[vertex,label=right:{$a_1^{\cycle}$}] (a1\cycle) at (\width,-\height) {};
        \node[vertex,label=left:{$b_1^{\cycle}$}] (b1\cycle) at (-\width,-\height) {};
        \draw[edge,draw=\cyclecolour]
            (a0\cycle) -- (b0\cycle) -- (a1\cycle) -- (b1\cycle) -- (a0\cycle);
    }

    \node[vertex,label=above:{$x_0$}] (x0) at (0,.95) {};
    \node[vertex,label=below:{$x_1$}] (x1) at (0,-.95) {};
    \node[vertex,label=above:{$y_0$}] (y0) at (0,1.775) {};
    \node[vertex,label=below:{$y_1$}] (y1) at (0,-1.775) {};
    \foreach \i in {0,1} {
        \foreach \letter in {a,b} {
            \draw[edge] (x\i) -- (\letter\i I);
            \draw[edge] (x\i) -- (\letter\i M);
            \draw[edge] (y\i) -- (\letter\i M);
            \draw[edge] (y\i) -- (\letter\i O);
        }
    }
\end{tikzpicture}
    \caption{A planar drawing of $H_1$.
    The highlighted cycles $I,M,O$ are the inner, middle, and outer rectangles, respectively.}
    \label{fig:grid-H1}
\end{figure}

We now define a planar graph $H_1$ that is an induced subgraph of a member of $\cB$ but is not an induced minor of any member of $\cT$; see \cref{fig:grid-H1}.
Take three disjoint $4$-cycles $I,M,O$ (\emph{inner}, \emph{middle}, and \emph{outer}), where the consecutive vertices of each cycle $C \in \{I,M,O\}$ are denoted by $a_0^C, b_0^C, a_1^C, b_1^C$.
We call the four vertices of $M$ the \emph{middle vertices}.
Add four pairwise nonadjacent vertices $x_0,x_1,y_0,y_1$ with
\[
    N(x_i)=\{a_i^I,b_i^I,a_i^M,b_i^M\},\qquad
    N(y_i)=\{a_i^M,b_i^M,a_i^O,b_i^O\}
    \quad \text{ for } i \in \{0,1\},
\]
and no other edges.
The graph $H_1$ has $16$ vertices.
\Cref{fig:grid-H1} shows that $H_1$ is planar.

An induced copy of $H_1$ in $B_{2,5}$ is given by mapping, for every $C \in \{I,M,O\}$ and $i \in \{0,1\}$,
\[
    a_i^C \mapsto (i+1,j_C,-),\qquad
    b_i^C \mapsto (i+1,j_C,+),\qquad
    x_i \mapsto (i+1,2,-),\qquad
    y_i \mapsto (i+1,4,-),
\]
where $j_I=1$, $j_M=3$, and $j_O=5$.
In particular, $H_1$ is an induced subgraph of every $B_{r,c}$ with $r \geq 2$ and $c \geq 5$.

To prove the exclusion from $\cT$, we need one further local property.

\begin{lemma}\label{lem:grid-side-agreement}
    Let $U,V$ be antiadjacent connected vertex sets in a graph from $\cT$, each with at least three vertices.
    If both sets meet a column, their union is not mixed in that column.
\end{lemma}
\begin{proof}    
    By \cref{lem:grid-barrier}\ref{item:grid-barrier-nonpath}, neither $U$ nor $V$ can be mixed in a column met by both sets: an opposite-side pair in one set would force the other to avoid that column.
    Suppose that $u \in U$ and $v \in V$ lie on opposite sides of a column $j$.
    Antiadjacency puts them in the same cell.
    Every horizontal neighbor of $u$ is therefore adjacent to $v$, so connectivity gives $u$ a vertical neighbor $u' \in U$.
    By \cref{lem:grid-barrier}\ref{item:grid-barrier-components}, the connected set $V$, which is antiadjacent to $\{u,u'\}$ and meets column $j$, has at most two vertices, a contradiction.
\end{proof}

\begin{lemma}\label{lem:grid-H1}
    No graph from $\cT$ contains $H_1$ as an induced minor.
\end{lemma}
\begin{proof}
    Suppose that $(X_v \mid v \in V(H_1))$ is an induced minor model of $H_1$ in $T_{r,c}$.
    We call the branch sets corresponding to middle vertices \emph{middle branch sets}.
    Let $U_I,U_M,U_O$ be the unions of the branch sets corresponding to the three cycles.
    These unions are connected, pairwise antiadjacent, and do not induce paths, since each contains a $4$-cycle as an induced minor.
    By \cref{lem:grid-barrier}\ref{item:grid-barrier-path}, each of these sets is mixed in some column.
    Choose columns $j_I,j_O$ in which $U_I,U_O$ are mixed, respectively.
    These columns are distinct by \cref{lem:grid-barrier}\ref{item:grid-barrier-nonpath}.
    By symmetry, we may assume $j_I<j_O$.
    Define
    \[
        J=\{j\in[c]\mid U_M\text{ is mixed in column }j\}
    \]
    and recall that every $j \in J$ is odd.
    The connected set $U_M\cup X_{y_0}\cup U_O$ is antiadjacent to $U_I$ and meets column $j_O>j_I$, so \cref{lem:grid-barrier}\ref{item:grid-barrier-nonpath} places it strictly to the right of $j_I$.
    Similarly, $U_I\cup X_{x_0}\cup U_M$ lies strictly to the left of $j_O$.
    Thus $U_M$ lies strictly between $j_I$ and $j_O$.
    For every $j\in J$, another application of \cref{lem:grid-barrier}\ref{item:grid-barrier-nonpath} places $U_I$ strictly to the left of $j$ and $U_O$ strictly to its right.

    Abbreviate $a_i=a_i^M$ and $b_i=b_i^M$, and put
    \[
        P_i^s=X_{x_i}\cup X_{s_i}\cup X_{y_i}
        \quad \text{ for } i \in \{0,1\},\ s \in \{a,b\}.
    \]
    For each $s\in\{a,b\}$, the sets $P_0^s,P_1^s$ are connected and antiadjacent, and each contains at least three vertices.
    By \cref{lem:grid-side-agreement}, their union is not mixed in any column they both meet.

    \begin{claim}\label{clm:grid-H1-paths}
        Each of the four middle branch sets induces a path contained in one row and meets every column $j \in J$.
        For every $j\in J$,
        \[
            X_{x_0},X_{x_1}\text{ lie strictly left of }j,\qquad
            X_{y_0},X_{y_1}\text{ lie strictly right of }j.
        \]
    \end{claim}
    \begin{claimproof}
        Fix $j\in J$.
        Since $U_I$ and $U_O$ lie on opposite sides of column $j$, and each $P_i^s$ has a neighbor in both, every $P_i^s$ meets column $j$.
        The sets $P_0^a,P_1^a$ use a common side in column $j$, and $P_0^b,P_1^b$ use a common side there.
        These sides must differ because $U_M$ is mixed in column $j$.
        For each $i\in\{0,1\}$, the branch sets $X_{x_i},X_{y_i}$ are contained in both $P_i^a$ and $P_i^b$, and hence cannot meet column $j$.
        Thus all four middle branch sets meet column $j$, with $X_{a_0},X_{a_1}$ on one side and $X_{b_0},X_{b_1}$ on the other.
        Connectivity and the adjacencies to the end cycles give the stated locations of $X_{x_0},X_{x_1},X_{y_0},X_{y_1}$.
        This argument applies to every $j\in J$.
        Thus no middle branch set is mixed in any column, and \cref{lem:grid-barrier}\ref{item:grid-barrier-path} makes each of them a path contained in one row.
    \end{claimproof}

    Fix $j\in J$.
    Choose $i$ so that the first column met by $X_{x_i}$ is no earlier than that met by $X_{x_{1-i}}$.
    By \cref{clm:grid-H1-paths}, every column met by $X_{x_i}$ is then met by $P_0^a,P_1^a,P_0^b,P_1^b$.
    Their side agreement forbids vertical edges within $X_{x_i}$ or from it to $X_{a_i}\cup X_{b_i}$.
    Since the middle branch sets induce paths in single rows, $X_{x_i},X_{a_i},X_{b_i}$ lie in one row, say $\rho$.

    Since $j$ is odd, there is a unique row-$\rho$ vertex $p_-$ in column $j-1$.
    At least one of the middle paths $X_{a_i},X_{b_i}$ avoids $p_-$, so its left endpoint lies in column $j$.
    Its adjacency to $X_{x_i}$ forces $p_-\in X_{x_i}$, and hence both middle paths avoid $p_-$.
    Similarly, the unique row-$\rho$ vertex in column $j+1$ belongs to $X_{y_i}$.
    Thus both middle paths are confined to cell $(\rho,j)$, whose two vertices are nonadjacent, a contradiction.
\end{proof}

\section{\NP-hardness of \textsc{Max Independent Set}}\label{sec:NPmis}
In this section we show that \MIS is \NP-hard in the hereditary closures of the families $\cB$ of loom grids and $\cT$ of triangle-free loom grids.
For $\cB$, this was already proved by Bonnet and Chang~\cite{BonnetChang2026}, but we include a self-contained (and slightly modified) proof for completeness.
The proof is a reduction from (unweighted) \textsc{Max Cut}, which asks whether a graph has a cut with at least a given number of edges. This problem is known to be \NP-hard~\cite{GAREY1976237}.

\paragraph{Intuition.}
To explain the intuition, consider first the graphs $B_{r,c}$ and $T_{r,c}$, where $c$ is odd.
Choosing all vertices on one side of each odd column and no vertices in even columns gives an independent set of size $r(c+1)/2$.
The side can be chosen independently in each odd column.
Assigning an odd column to each input vertex therefore lets us encode the side of a cut by these choices of sides.

In the reduction, we retain suitable vertices in two rows for each input edge.
After choosing a side in every odd column, the available vertices in these rows induce disjoint paths consisting only of horizontal edges.
The maximum contribution of the two rows assigned to an input edge is a fixed amount, plus one exactly when its endpoints are assigned different sides.
The construction ensures that a maximum independent set can be obtained from such a choice of sides, so maximizing the independent set maximizes the cut.

\paragraph{The reduction.}
Let $F$ be a graph with ordered vertices $v_1,\ldots,v_n$, where $n\geq 1$, and $m$ edges.
Write $\mc(F)$ for the maximum size of a cut in $F$, i.e., the maximum number of bicolored edges in any 2-coloring of $V(F)$.
By $\alpha(G)$ we denote the maximum size of an independent set in a graph $G$.

We use $2m+2$ rows and $2n-1$ columns.
The first two rows play a special role and are called \emph{guards}; we denote them by $g^-$ and $g^+$.
The remaining $2m$ rows correspond to the edges of $F$, with two rows for each edge.
For each edge $e\in E(F)$, we label its two rows $e^-$ and $e^+$.
Using these labels as row coordinates, we construct induced subgraphs $G_B=G_B(F)$ of $B_{2m+2,2n-1}$ and $G_T=G_T(F)$ of $T_{2m+2,2n-1}$ by retaining the same vertex set in both hosts.
For a side $s\in\{-,+\}$, let $\bar s$ denote the opposite side.
For every edge $e=v_av_b$, where $a<b$, and each $s\in\{-,+\}$, retain in row $e^s$ the vertices
\[
    (e^s,2a-1,s),\qquad (e^s,2b-1,\bar s),
\]
as well as both vertices in every other odd column, and the side-$-$ vertex in every even column.
In the guard rows retain only the vertices
\[
    g_j^s=(g^s,2j-1,s)\qquad \text{ for } j\in[n],\ s\in\{-,+\}.
\]
We call these vertices \emph{guard vertices}.
Selecting $g_j^s$ in an independent set forces every selected vertex in column $2j-1$ to lie on side $s$.
The graph $G_T$ is obtained from $G_B$ by deleting all internal edges.
Each guard vertex is the only retained vertex in its cell, so its adjacencies are the same in both graphs.
All edges incident with guard vertices are vertical: each pair $g_j^-,g_j^+$ is adjacent, and $g_j^s$ is adjacent to every retained vertex on the opposite side of column $2j-1$ and to no vertex outside that column.
Each edge row has $2n-2$ vertices in odd columns and $n-1$ in even columns, hence $3(n-1)$ vertices in total, so
\[
    |V(G_B)|=|V(G_T)|=6m(n-1)+2n=\Oh(nm+n).
\]
In particular, both graphs can be constructed in polynomial time.

\begin{lemma}\label{lem:mwis-encoding}
    Let $F$ be an $n$-vertex $m$-edge graph and let $G_B=G_B(F)$ and $G_T=G_T(F)$. Then
    \[
        \alpha(G_B)=\alpha(G_T)=n+2m(n-1)+\mc(F).
    \]
\end{lemma}
\begin{proof}
    Fix $G\in\{G_B,G_T\}$.
    Call an independent set of $G$ \emph{canonical} if it contains exactly one guard vertex in every odd column.

    \begin{claim}\label{clm:mwis-guards}
        There is a canonical maximum independent set of $G$.
    \end{claim}
    \begin{claimproof}
        The two guard vertices in each odd column are adjacent, so an independent set contains at most one of them.
        Among all maximum independent sets of $G$, choose $I$ containing the largest number of guard vertices.
        Suppose that $I$ is not canonical, so some odd column $j$ contains no guard vertex of $I$.
        If $I$ is not mixed in column $j$, we can add a guard vertex on its side, choosing either side when the column is empty, contradicting the maximality of $I$.
        Thus $I$ is mixed in column $j$, and $G=G_T$, since in $G_B$ any opposite-side pair is joined by a vertical or internal edge.
        In $G_T$, opposite-side vertices in different rows are adjacent, so the vertices of $I$ in column $j$ are exactly the two vertices of one cell.
        Replace one by the guard vertex on the side of the other to obtain $I'$.
        Since guard vertices have no neighbors outside their columns, $I'$ is an independent set of size $|I|$ with one more guard vertex than $I$, contradicting the choice of $I$.
    \end{claimproof}

    Each canonical independent set $I$ defines a cut of $F$ with parts $\{v_j\mid g_j^-\in I\}$ and $\{v_j\mid g_j^+\in I\}$.
    For each $j\in[n]$, let $s_j$ be the unique side such that $g_j^{s_j}\in I$; thus $v_av_b$ crosses this cut exactly when $s_a\ne s_b$.

    \begin{claim}\label{clm:mwis-cut}
        If a canonical independent set $I$ defines a cut with $k$ crossing edges, then
        \[
            |I|\leq n+2m(n-1)+k.
        \]
    \end{claim}
    \begin{claimproof}
        Since $I$ is canonical, each edge row contains at most one selected vertex per column, and its selected columns are nonconsecutive.
        Thus each row contributes at most $n$ vertices, with equality only if every odd column is represented.
        For $e=v_av_b$, where $a<b$, equality in row $e^s$ requires $s=s_a$ and $\bar s=s_b$.
        Hence at most one of its two rows contributes $n$ vertices if $e$ crosses the cut, and neither can otherwise.
        Their total contribution is therefore at most $2(n-1)$, plus one for a crossing edge.
        Summing and adding the $n$ guard vertices proves the bound.
    \end{claimproof}

    \begin{claim}\label{clm:mwis-cut-realization}
        For every cut of $F$ with $k$ crossing edges, there is a canonical independent set $I$ defining that cut and satisfying
        \[
            |I|=n+2m(n-1)+k.
        \]
    \end{claim}
    \begin{claimproof}
        Fix a cut with parts indexed by $-$ and $+$, and let $s_j$ be the side containing $v_j$.
        Select the guard vertices $g_j^{s_j}$.
        For each crossing edge $e=v_av_b$, where $a<b$, select $(e^{s_a},2j-1,s_j)$ for every $j\in[n]$; these vertices exist because $s_b=\overline{s_a}$.
        In all remaining edge rows, including the second row of each crossing edge, select all even-column vertices.

        All selected vertices in each column lie on the same side, and no row uses consecutive columns.
        Thus the resulting set is independent, canonical, and defines the given cut.
        Exactly $k$ edge rows contribute $n$ vertices, and the remaining $2m-k$ contribute $n-1$, giving the required size after adding the $n$ guards.
    \end{claimproof}

    By \cref{clm:mwis-guards,clm:mwis-cut,clm:mwis-cut-realization}, maximizing over the cuts of $F$ gives
    $\alpha(G)=n+2m(n-1)+\mc(F)$, which completes the proof of the lemma.
\end{proof}

Given an instance $(F,k)$ of \textsc{Max Cut}, consider an instance $(G,k')$ of \MIS, where $G$ is either $G_B$ or $G_T$ and $k' = n+2m(n-1)+k$.
By \cref{lem:mwis-encoding}, the resulting instance is a yes-instance if and only if $F$ has a cut of size at least $k$.
The construction is polynomial and produces a graph in the hereditary closure of $\cB$ or $\cT$, respectively, yielding the following two results.

\thmmwisB*

\thmmwisT*

\section{\NP-hardness of \col{3}}\label{sec:coloring}\label{sec:color}
In this section, we show that \textsc{3-Coloring} is NP-hard in the hereditary closure of the family $\mathcal{B}$ of loom grids and in the induced minor closure of the family $\mathcal{T}$ of triangle-free loom grids.
Note that every triangle-free loom grid, and hence every graph in the hereditary closure of $\mathcal{T}$, is $3$-colorable.
We give a common reduction from the \textsc{3-Sat} problem which was shown to be \NP-hard in~\cite{Cook1971}.
As in \cref{sec:NPmis}, a choice of one side in each of certain columns encodes a solution of the input instance.
Here these choices encode the truth values of the variables while each clause is defined by its own row.
Along each clause row, a sequence of state vertices uses two reference colors and is forced to start and end with different colors.
Four additional reference rows ensure that every proper $3$-coloring determines such choices consistently.

\paragraph{The reduction.}
Let $\Phi$ be a CNF formula with at most three literals per clause.
First, delete tautological clauses and repeated occurrences of a literal within a clause.
Clearly, these operations preserve satisfiability.
We write $x_1,\ldots,x_n$ for the variables and $C_1,\ldots,C_m$ for the remaining clauses.
We may assume that $n\ge1$.
We set $r=m+4$ and $t=2n+1$.
The set of row labels is
\[
 R\coloneqq [m]\cup\{a_1,a_2,b_1,b_2\},
 \qquad \mathsf{A}=\{a_1,a_2\},\qquad \mathsf{B}=\{b_1,b_2\}.
\]
The rows in $\mathsf{A}\cup\mathsf{B}$ are called the \emph{reference rows}, while each row $i\in[m]$ corresponds to $C_i$.

We construct two graphs $G_B=G_B(\Phi)$ and $G_T=G_T(\Phi)$ on the same vertex set.
For every row $i\in R$, we introduce \emph{state vertices} $p_{i,0},\ldots,p_{i,t}$ and \emph{cells} $D_{i,1},\ldots,D_{i,t}$, where each cell consists of some of the two vertices $d_{i,j}^{-},d_{i,j}^{+}$.
A cell is \emph{full} if both vertices are present.
All reference cells are full.
In each clause row, every odd-indexed cell is a full cell and is called an \emph{auxiliary cell}, while the variables are represented by \emph{variable cells} defined as follows:
\begin{equation}\label{eq:col-clause-cells}
 D_{i,2k}=
 \begin{cases}
  \{d_{i,2k}^{-}\},&\text{if }x_k\in C_i,\\
  \{d_{i,2k}^{+}\},&\text{if }\neg x_k\in C_i,\\
  \{d_{i,2k}^{-},d_{i,2k}^{+}\},&\text{if neither literal occurs in }C_i.
 \end{cases}
\end{equation}
The index $j\in[t]$ is called the \emph{cell column}.
It is distinct from the column coordinate in a host loom grid.

\begin{figure}[htbp]
\centering
% This picture uses only the basic TikZ package.
% The example is Phi = (x_1 or not x_2) and (not x_1), with r=6 and t=5.
% Plot rows 1,...,4 are a_1,a_2,b_1,b_2, and plot rows 5,6 are the clause rows.
% Every edge of G_B is drawn, and deleting the red internal edges gives G_T.
\begin{tikzpicture}[
    x=1cm,y=1cm,
    vertex/.style={circle,draw=black,fill=white,inner sep=0pt,minimum size=4pt,line width=.65pt},
    state vertex/.style={vertex,fill=black!20},
    vertical edge/.style={draw=blue!65!black,line width=.45pt,opacity=.7},
    horizontal edge/.style={draw=black,line width=.55pt},
    internal edge/.style={draw=red!75!black,line width=.65pt},
    every node/.style={font=\small,inner sep=2pt},
]
    % The only omitted vertices are d^+_{1,2}, d^-_{1,4}, and d^-_{2,2}.
    \def\colfigcelltype#1#2{%
        \pgfmathtruncatemacro{\hasminus}{!(((#1==5)&&(#2==4))||((#1==6)&&(#2==2)))}%
        \pgfmathtruncatemacro{\hasplus}{!((#1==5)&&(#2==2))}%
    }

    % Define all positions before drawing the edges.
    \foreach \plotrow in {1,...,6} {
        \pgfmathsetmacro{\rowy}{-.88*(\plotrow-1)-.22*(\plotrow>4)}
        \foreach \column in {1,...,5} {
            \coordinate (c-\plotrow-\column-minus) at
                ({2.2*(\column-.5)-.18},{\rowy+.10});
            \coordinate (c-\plotrow-\column-plus) at
                ({2.2*(\column-.5)+.18},{\rowy-.10});
        }
        \foreach \state in {0,...,5} {
            \ifnum\state=0\relax
                \pgfmathtruncatemacro{\stateside}{ifthenelse(\plotrow<=2,-1,1)}
                \coordinate (p-\plotrow-\state) at
                    ({.18*\stateside},{\rowy-.10*\stateside});
            \else
                \ifnum\state=5\relax
                    \pgfmathtruncatemacro{\stateside}{ifthenelse((\plotrow==3)||(\plotrow==4),1,-1)}
                    \coordinate (p-\plotrow-\state) at
                        ({11+.18*\stateside},{\rowy-.10*\stateside});
                \else
                    \coordinate (p-\plotrow-\state) at ({2.2*\state},\rowy);
                \fi
            \fi
        }
    }

    % Cell-column edges join surviving opposite-side vertices in distinct rows.
    \foreach \column in {1,...,5} {
        \foreach \plotrow in {1,...,6} {
            \colfigcelltype{\plotrow}{\column}
            \ifnum\hasminus=1\relax
                \foreach \otherrow in {1,...,6} {
                    \colfigcelltype{\otherrow}{\column}
                    \ifnum\hasplus=1\relax
                        \ifnum\plotrow=\otherrow\relax\else
                            \draw[vertical edge] (c-\plotrow-\column-minus)
                                -- (c-\otherrow-\column-plus);
                        \fi
                    \fi
                }
            \fi
        }
    }

    % The two boundary bicliques are K_{2,4}.
    \foreach \arow in {1,2} {
        \foreach \otherrow in {3,...,6}
            \draw[vertical edge] (p-\arow-0) -- (p-\otherrow-0);
    }
    \foreach \brow in {3,4} {
        \foreach \otherrow in {1,2,5,6}
            \draw[vertical edge] (p-\brow-5) -- (p-\otherrow-5);
    }

    % Every surviving cell vertex sees the two neighboring state vertices.
    \foreach \plotrow in {1,...,6} {
        \foreach \column in {1,...,5} {
            \pgfmathtruncatemacro{\previousstate}{\column-1}
            \colfigcelltype{\plotrow}{\column}
            \ifnum\hasminus=1\relax
                \draw[horizontal edge] (p-\plotrow-\previousstate)
                    -- (c-\plotrow-\column-minus) -- (p-\plotrow-\column);
            \fi
            \ifnum\hasplus=1\relax
                \draw[horizontal edge] (p-\plotrow-\previousstate)
                    -- (c-\plotrow-\column-plus) -- (p-\plotrow-\column);
            \fi
        }
    }

    % The internal edges distinguish G_B from G_T.
    \foreach \plotrow in {1,...,6} {
        \foreach \column in {1,...,5} {
            \colfigcelltype{\plotrow}{\column}
            \ifnum\hasminus=1\relax
                \ifnum\hasplus=1\relax
                    \draw[internal edge] (c-\plotrow-\column-minus)
                        -- (c-\plotrow-\column-plus);
                \fi
            \fi
        }
    }

    % Draw vertices last so that edges do not pass through their interiors.
    \foreach \plotrow/\rowlabel in {1/a_1,2/a_2,3/b_1,4/b_2,5/C_1,6/C_2} {
        \pgfmathsetmacro{\rowy}{-.88*(\plotrow-1)-.22*(\plotrow>4)}
        \node at (-.7,\rowy) {$\rowlabel$};
        \foreach \state in {0,...,5}
            \node[state vertex] at (p-\plotrow-\state) {};
        \foreach \column in {1,...,5} {
            \colfigcelltype{\plotrow}{\column}
            \ifnum\hasminus=1\relax
                \node[vertex] at (c-\plotrow-\column-minus) {};
            \fi
            \ifnum\hasplus=1\relax
                \node[vertex] at (c-\plotrow-\column-plus) {};
            \fi
        }
    }

    % The headings use cell-column indices, not host-grid coordinates.
    \node at (5.5,1.65) {$\Phi=(x_1\vee\neg x_2)\wedge(\neg x_1)$};
    \node[font=\scriptsize] at (0,.8) {$p_{i,0}$};
    \node[font=\scriptsize] at (11,.8) {$p_{i,5}$};
    \foreach \column/\columnlabel in {1/{\mathrm{aux.}},2/x_1,3/{\mathrm{aux.}},4/x_2,5/{\mathrm{aux.}}} {
        \node at ({2.2*(\column-.5)},1.1) {$\column$};
        \node[font=\scriptsize] at ({2.2*(\column-.5)},.8) {$\columnlabel$};
        \node[font=\scriptsize] at ({2.2*(\column-.5)-.18},.43) {$-$};
        \node[font=\scriptsize] at ({2.2*(\column-.5)+.18},.43) {$+$};
    }

    % The legend distinguishes state vertices and the removable internal edges.
    \node[state vertex] at (0,-5.35) {};
    \node[anchor=west,font=\scriptsize] at (.15,-5.35) {state vertex};
    \node[vertex] at (3.2,-5.35) {};
    \node[anchor=west,font=\scriptsize] at (3.35,-5.35) {cell vertex};
    \draw[internal edge] (6.3,-5.35) -- (6.8,-5.35);
    \node[anchor=west,font=\scriptsize] at (6.95,-5.35) {internal edge ($G_B$ only)};
\end{tikzpicture}
\caption{The construction for $\Phi=(x_1\vee\neg x_2)\wedge(\neg x_1)$.
All edges of $G_B(\Phi)$ are shown; deleting the red internal edges gives $G_T(\Phi)$.
Gray vertices are state vertices; black edges join cell vertices to their neighboring states, and blue edges join vertices in different rows.
The first four rows are the reference rows, and the indices $j=1,\ldots,5$ denote cell columns.}
\label{fig:col-construction}
\end{figure}
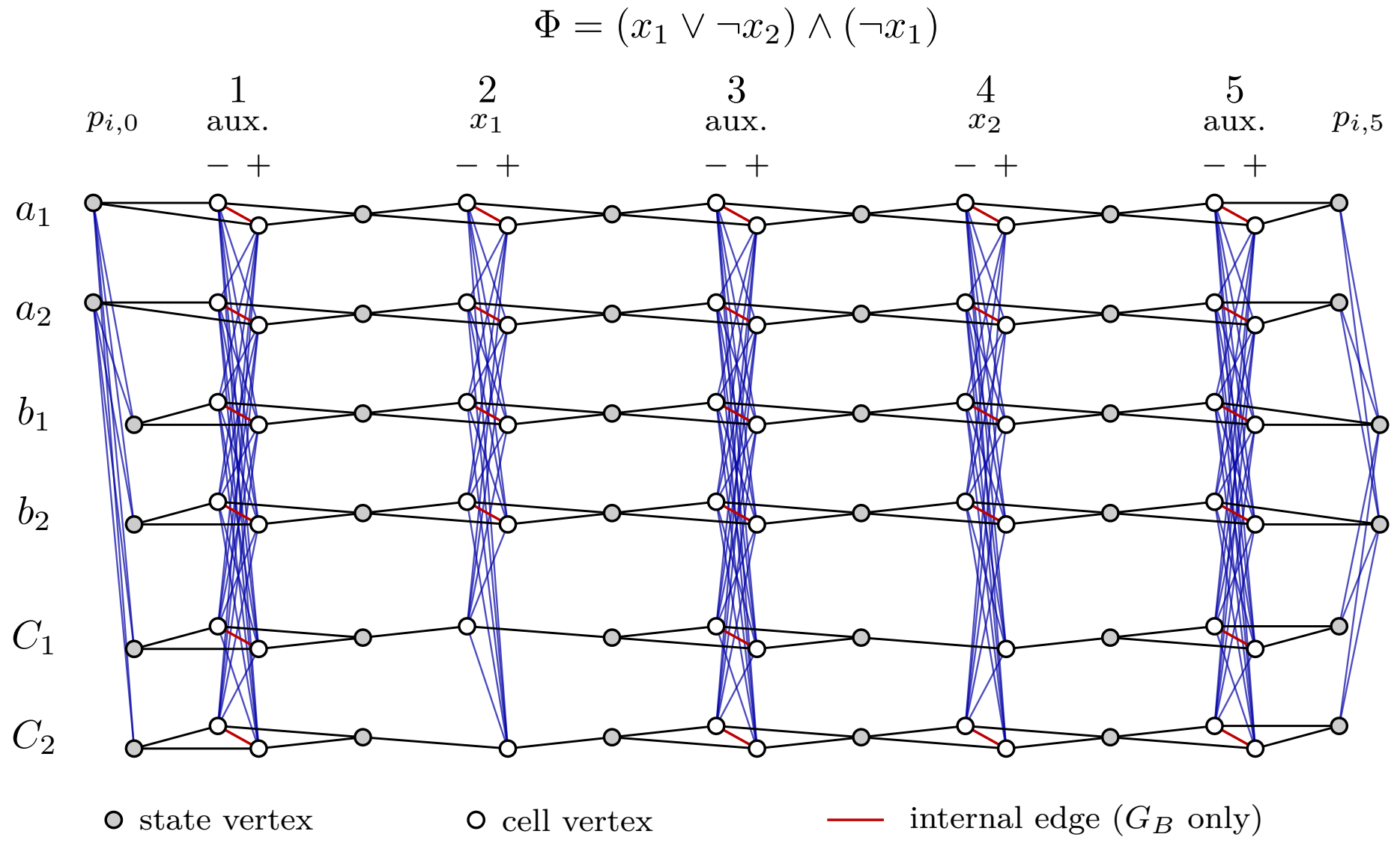

In both graphs, each vertex of $D_{i,j}$ is adjacent to $p_{i,j-1}$ and $p_{i,j}$.
Within each cell column, join every present $-$-side vertex to every present $+$-side vertex in a different row.
In $G_B$, also add the internal edge $d_{i,j}^{-}d_{i,j}^{+}$ in every full cell.
These edges are absent in $G_T$.
At the left boundary, make $\{p_{a,0}:a\in\mathsf{A}\}$ complete to $\{p_{i,0}:i\in R\setminus\mathsf{A}\}$.
At the right boundary, make $\{p_{b,t}:b\in\mathsf{B}\}$ complete to $\{p_{i,t}:i\in R\setminus\mathsf{B}\}$.
There are no further edges.
Thus $G_T$ is a spanning subgraph of $G_B$, obtained by deleting all internal cell edges.
See \cref{fig:col-construction} for an example of the construction.

Each row has $t+1$ state vertices and at most $2t$ cell vertices, so
\begin{equation}\label{eq:col-size}
 |V(G_B)|=|V(G_T)|\leq (m+4)(3t+1)=(m+4)(6n+4).
\end{equation}
Both graphs can therefore be constructed in polynomial time.

\pagebreak[2]

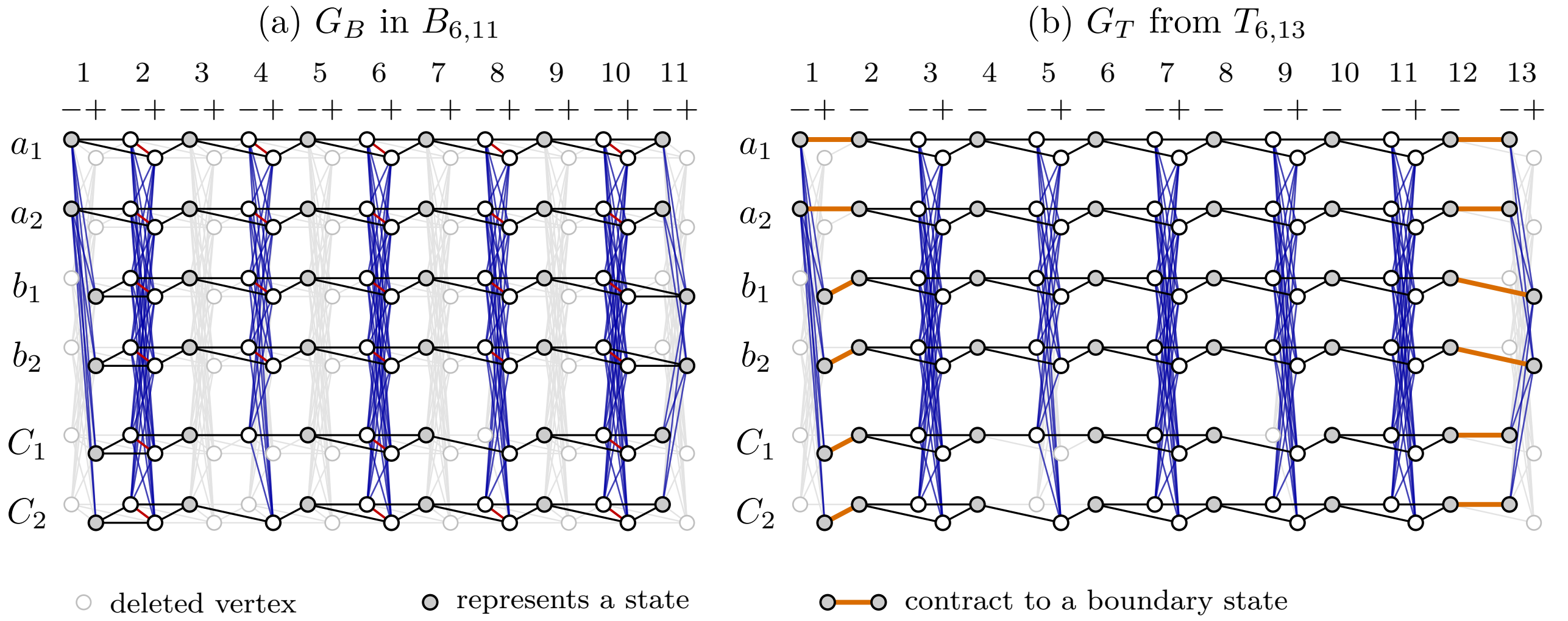
\begin{figure}[t]
    \centering
    \resizebox{\linewidth}{!}{% This picture uses only the basic TikZ package.
% It illustrates Lemma 5.1 for Phi = (x_1 or not x_2) and (not x_1), with r=6 and t=5.
% The visual row order is a_1, a_2, b_1, b_2, C_1, C_2.
% Delete the pale vertices in both panels, and contract the orange edges in the right panel.
% Host-column numbers and side labels are retained, while repeated construction-role headings are omitted.
\begin{tikzpicture}[
    x=1cm,y=1cm,
    vertex/.style={circle,draw=black,fill=white,inner sep=0pt,minimum size=4pt,line width=.65pt},
    state vertex/.style={vertex,fill=black!20},
    deleted vertex/.style={vertex,draw=black!25,line width=.45pt},
    deleted edge/.style={draw=black!11,line width=.40pt},
    vertical edge/.style={draw=blue!65!black,line width=.45pt,opacity=.7},
    horizontal edge/.style={draw=black,line width=.55pt},
    internal edge/.style={draw=red!75!black,line width=.65pt},
    contract edge/.style={draw=orange!85!black,line width=1.35pt},
    every node/.style={font=\small,inner sep=2pt},
]
    % A side value of 0 means minus, and a side value of 1 means plus.
    % The parameter kind is 0 for B and 1 for T.
    \def\hostfigkeep#1#2#3{%
        \pgfmathtruncatemacro{\keepthis}{%
            ifthenelse(#2==1,
                ifthenelse(#1<=2,#3==0,#3==1),
                ifthenelse(#2==\lastcolumn,
                    ifthenelse((#1==3)||(#1==4),#3==1,#3==0),
                    ifthenelse(mod(#2,2)==\kind,
                        !(((#1==5)&&(#2==4+\kind)&&(#3==1))
                          ||((#1==5)&&(#2==8+\kind)&&(#3==0))
                          ||((#1==6)&&(#2==4+\kind)&&(#3==0))),
                        #3==0)))}%
    }

    % Draw the whole host faintly on pass 0 and its retained edges on pass 1.
    \def\hostfigedge#1#2#3#4#5#6#7{%
        \ifnum\drawpass=0\relax
            \draw[deleted edge] (g-\kind-#1-#2-#3) -- (g-\kind-#4-#5-#6);
        \else
            \hostfigkeep{#1}{#2}{#3}%
            \let\keepfirst\keepthis
            \hostfigkeep{#4}{#5}{#6}%
            \ifnum\keepfirst=1\relax
                \ifnum\keepthis=1\relax
                    \draw[#7] (g-\kind-#1-#2-#3) -- (g-\kind-#4-#5-#6);
                \fi
            \fi
        \fi
    }

    \node at (6.8,1.85) {$\Phi=(x_1\vee\neg x_2)\wedge(\neg x_1)$};
    \foreach \kind/\lastcolumn/\panelx/\panely in {0/11/0/0,1/13/7.15/0} {
        \begin{scope}[xshift=\panelx cm,yshift=\panely cm]
            % Even columns of T have no plus vertex, including in the pale host.
            \foreach \column in {1,...,\lastcolumn} {
                \pgfmathtruncatemacro{\lastside}{(\kind==0)||(mod(\column,2)==1)}
                \foreach \row in {1,...,6} {
                    \pgfmathsetmacro{\rowy}{-.68*(\row-1)-.18*(\row>4)}
                    \foreach \side in {0,...,\lastside}
                        \coordinate (g-\kind-\row-\column-\side) at
                            ({.58*(\column-1)+.12*(2*\side-1)},
                             {\rowy-.09*(2*\side-1)});
                }
            }

            \foreach \drawpass in {0,1} {
                % Opposite sides in distinct rows give the vertical edges.
                \foreach \column in {1,...,\lastcolumn} {
                    \pgfmathtruncatemacro{\hasplus}{(\kind==0)||(mod(\column,2)==1)}
                    \ifnum\hasplus=1\relax
                        \foreach \row in {1,...,6} {
                            \foreach \otherrow in {1,...,6} {
                                \ifnum\row=\otherrow\relax\else
                                    \hostfigedge{\row}{\column}{0}{\otherrow}{\column}{1}{vertical edge}
                                \fi
                            }
                        }
                    \fi
                }

                % Consecutive cells in a row are complete to each other.
                \pgfmathtruncatemacro{\penultimatecolumn}{\lastcolumn-1}
                \foreach \column in {1,...,\penultimatecolumn} {
                    \pgfmathtruncatemacro{\nextcolumn}{\column+1}
                    \pgfmathtruncatemacro{\sourcesides}{(\kind==0)||(mod(\column,2)==1)}
                    \pgfmathtruncatemacro{\targetsides}{(\kind==0)||(mod(\nextcolumn,2)==1)}
                    \def\edgestyle{horizontal edge}
                    \ifnum\kind=1\relax
                        \ifnum\column=1\relax
                            \def\edgestyle{contract edge}
                        \fi
                        \ifnum\column=12\relax
                            \def\edgestyle{contract edge}
                        \fi
                    \fi
                    \foreach \row in {1,...,6} {
                        \foreach \side in {0,...,\sourcesides} {
                            \foreach \otherside in {0,...,\targetsides} {
                                \hostfigedge{\row}{\column}{\side}{\row}{\nextcolumn}{\otherside}{\edgestyle}
                            }
                        }
                    }
                }

                % Only B contains the internal edges of its cells.
                \ifnum\kind=0\relax
                    \foreach \column in {1,...,\lastcolumn} {
                        \foreach \row in {1,...,6} {
                            \hostfigedge{\row}{\column}{0}{\row}{\column}{1}{internal edge}
                        }
                    }
                \fi
            }

            % Retained state representatives are gray, and retained cell vertices are white.
            \foreach \column in {1,...,\lastcolumn} {
                \pgfmathtruncatemacro{\lastside}{(\kind==0)||(mod(\column,2)==1)}
                \pgfmathtruncatemacro{\isstate}{(\column==1)||(\column==\lastcolumn)||(mod(\column,2)!=\kind)}
                \foreach \row in {1,...,6} {
                    \foreach \side in {0,...,\lastside} {
                        \hostfigkeep{\row}{\column}{\side}
                        \ifnum\keepthis=0\relax
                            \node[deleted vertex] at (g-\kind-\row-\column-\side) {};
                        \else
                            \ifnum\isstate=1\relax
                                \node[state vertex] at (g-\kind-\row-\column-\side) {};
                            \else
                                \node[vertex] at (g-\kind-\row-\column-\side) {};
                            \fi
                        \fi
                    }
                }
            }

            % Number every host column and indicate its available sides.
            \foreach \column in {1,...,\lastcolumn} {
                \node[font=\scriptsize] at ({.58*(\column-1)},.75) {$\column$};
                \node[font=\scriptsize] at ({.58*(\column-1)-.12},.38) {$-$};
                \pgfmathtruncatemacro{\hasplus}{(\kind==0)||(mod(\column,2)==1)}
                \ifnum\hasplus=1\relax
                    \node[font=\scriptsize] at ({.58*(\column-1)+.12},.38) {$+$};
                \fi
            }
            \foreach \row/\rowlabel in {1/a_1,2/a_2,3/b_1,4/b_2,5/C_1,6/C_2}
                \node at (-.55,{-.68*(\row-1)-.18*(\row>4)}) {$\rowlabel$};

            \ifnum\kind=0\relax
                \node at (2.9,1.22) {(a) $G_B$ in $B_{6,11}$};
            \else
                \node at (3.48,1.22) {(b) $G_T$ from $T_{6,13}$};
            \fi
        \end{scope}
    }

    % The legend is shared by both panels.
    \node[deleted vertex] at (0,-4.45) {};
    \node[anchor=west,font=\scriptsize] at (.18,-4.45) {deleted vertex};
    \node[state vertex] at (3.4,-4.45) {};
    \node[anchor=west,font=\scriptsize] at (3.58,-4.45) {represents a state};
    \draw[contract edge] (7.3,-4.45) -- (7.8,-4.45);
    \node[state vertex] at (7.3,-4.45) {};
    \node[state vertex] at (7.8,-4.45) {};
    \node[anchor=west,font=\scriptsize] at (7.98,-4.45) {contract to a boundary state};
\end{tikzpicture}}
    \caption{The host models from Lemma~\ref{lem:col-hosts} for $\Phi=(x_1\vee\neg x_2)\wedge(\neg x_1)$.
    In the left panel, deleting the pale vertices yields an induced copy of $G_B$ in $B_{6,11}$.
    In the right panel, deleting the pale vertices and contracting the twelve orange boundary edges yields $G_T$ from $T_{6,13}$.
    Gray vertices represent state vertices or belong to their two-vertex boundary branch sets.
    Blue edges are vertical, black edges are horizontal, and red edges are internal; the orange edges are the horizontal edges to be contracted.
    The column numbers refer to the host loom grids.}
    \label{fig:col-hosts}
\end{figure}

\begin{lemma}\label{lem:col-hosts}
The graph $G_B$ is an induced subgraph of $B_{r,2t+1}$, and $G_T$ is an induced minor of $T_{r,2t+3}$.
\end{lemma}

The two models are illustrated in \cref{fig:col-hosts} for the example from \cref{fig:col-construction}.
In the left panel, deleting the pale vertices of the host loom grid leaves exactly $G_B$.
In the right panel, deleting the pale vertices and contracting the orange edges gives $G_T$, with each contracted edge becoming a boundary state vertex.

\begin{proof}[Proof of \cref{lem:col-hosts}]
Use the labels in $R$ as row coordinates and, for $i\in R$, put
\[
 \lambda_i=\begin{cases}-,&i\in\mathsf{A},\\+,&i\notin\mathsf{A},\end{cases}
 \qquad
 \rho_i=\begin{cases}+,&i\in\mathsf{B},\\-,&i\notin\mathsf{B}.\end{cases}
\]
In $B_{r,2t+1}$, represent each present cell vertex $d_{i,j}^{s}$ by $(i,2j,s)$, and represent the state vertices by
\[
 p_{i,0}\mapsto(i,1,\lambda_i),\qquad
 p_{i,j}\mapsto(i,2j+1,-)\quad(0<j<t),\qquad
 p_{i,t}\mapsto(i,2t+1,\rho_i).
\]
The interior state columns are independent, and the boundary columns induce the prescribed bicliques.
All other edges are exactly the cell-column and cell--state edges, so this is an induced copy of $G_B$.

In $T_{r,2t+3}$, use the following branch sets, omitting any absent cell vertices:
\begin{align*}
 X_{d_{i,j}^{s}}&=\{(i,2j+1,s)\} &&(j\in[t]),\\
 X_{p_{i,j}}&=\{(i,2j+2,-)\} &&(0<j<t),\\
 X_{p_{i,0}}&=\{(i,1,\lambda_i),(i,2,-)\},\\
 X_{p_{i,t}}&=\{(i,2t+2,-),(i,2t+3,\rho_i)\}.
\end{align*}
These sets are pairwise disjoint and connected.
The odd-column vertices in the boundary branch sets induce the required bicliques, since only one vertex per row is retained in each boundary column.
The even-column vertex in each boundary branch set supplies the adjacencies to the first or last cell in its row.
All other adjacencies are exactly the cell-column and cell--state adjacencies, so this is an induced minor model of $G_T$.
\end{proof}

\begin{lemma}\label{lem:col-correctness}
For every formula $\Phi$ as above, the following are equivalent: $\Phi$ is satisfiable, $G_B(\Phi)$ is $3$-colorable, and $G_T(\Phi)$ is $3$-colorable.
\end{lemma}
\begin{proof}
Since $G_T$ is a spanning subgraph of $G_B$, it suffices to show that a proper $3$-coloring of $G_T$ gives a satisfying assignment of $\Phi$, and that a satisfying assignment gives a proper $3$-coloring of $G_B$.

\paragraph{The reference rows.}
Fix a proper coloring $\varphi:V(G_T)\to\{1,2,3\}$.
We show by induction over $j\in[t]$ that every reference cell $D_{i,j}$ uses two colors and
\[
 \varphi(p_{i,j})=\varphi(p_{i,j-1})=\varphi(p_{i,0})
 \qquad\text{for every }i\in\mathsf{A}\cup\mathsf{B}.
\]
The initial reference states induce a $K_{2,2}$, so no three have the same color.
By the induction hypothesis, the same holds for the states preceding cell column $j$.
If one reference cell were monochromatic with color $c$, the other three reference cells would induce a $6$-cycle colored with the other two colors.
The bipartition of this cycle is given by the sides, so each of these three cells would use both remaining colors.
Their preceding states would therefore all have color $c$, a contradiction.
Thus every reference cell uses two colors, forcing its two neighboring states to have the same, remaining color.
This completes the induction.

\paragraph{From the coloring to an assignment.}
Relabel the colors so that the states in row $a_1$ have color $1$ and those in row $b_1$ have color $2$; their initial states are adjacent.
In each cell column $j$, the cell in row $a_1$ uses colors $2,3$, and the cell in row $b_1$ uses colors $1,3$.
Their color-$3$ vertices lie on the same side, since otherwise they would be adjacent.
Denote this side by $\sigma_j$ and call it the \emph{selected side}.
Every cell vertex in another row on side $\sigma_j$ sees reference vertices of colors $1$ and $2$, and therefore has color~$3$.
Every cell vertex on the opposite side sees a reference vertex of color $3$, and therefore has color $1$ or $2$.
Thus, in every row, cell vertices on the selected side have color $3$ and those on the opposite side have color $1$ or $2$.

Every clause state is adjacent to a full auxiliary cell and therefore has color $1$ or $2$.
The boundary edges to the states in rows $a_1$ and $b_1$ then force
\[
 \varphi(p_{i,0})=2,\qquad \varphi(p_{i,t})=1
 \qquad\text{for every }i\in[m].
\]
If a cell in a clause row contains a vertex on the nonselected side, that vertex and both neighboring states use colors from $\{1,2\}$, so the neighboring states have the same color.
Thus a change between consecutive state colors is possible only at a singleton on the selected side.

Set $x_k$ to true if and only if $\sigma_{2k}=-$.
By \eqref{eq:col-clause-cells}, a singleton on the selected side represents a true literal.
Every clause row must change from color $2$ to color $1$ somewhere, so every clause contains a true literal.
Hence $\Phi$ is satisfiable.

\paragraph{From an assignment to a coloring.}
Conversely, fix a satisfying assignment of $\Phi$.
In cell column $2k$, select side $-$ if $x_k$ is true and side $+$ otherwise.
In every odd cell column, select either side.
Give all states in the rows of $\mathsf{A}$ color $1$ and all states in the rows of $\mathsf{B}$ color $2$.
In each clause $C_i$, choose one true literal, and let $j_i$ be the index of its cell column.
Color $p_{i,j}$ with $2$ for $j<j_i$ and with $1$ for $j\geq j_i$.

Give every cell vertex on the selected side color $3$.
Every remaining cell vertex has two equally colored neighboring states, since a change occurs only at a selected-side singleton.
Color each remaining vertex with the other color in $\{1,2\}$.
All cell--state edges are properly colored.
Every edge within a cell column, including each internal edge of $G_B$, joins color $3$ to color $1$ or $2$, and every boundary edge joins colors $1$ and $2$.
This is a proper $3$-coloring of $G_B$.
\end{proof}

By Lemma~\ref{lem:col-correctness}, the two constructions are reductions from \textsc{3-Sat} to \textsc{3-Coloring}.
They are polynomial by \eqref{eq:col-size}, and Lemma~\ref{lem:col-hosts} places their outputs in the required graph classes.
Since \textsc{3-Sat} is NP-complete, we obtain the following two results.

\thmcolB*

\thmcolT*

\section{Dominated balanced separators}\label{sec:dbs}
Let $H_{1,1}$ be the six-vertex tree with four leaves and two vertices
of degree $3$. Denote its vertices as follows:
the two vertices of degree $3$ are $b$ and $q$, the two
leaves neighboring $b$ are $a$ and $c$ and the two leaves neighboring
$q$ are $p$ and $r$.

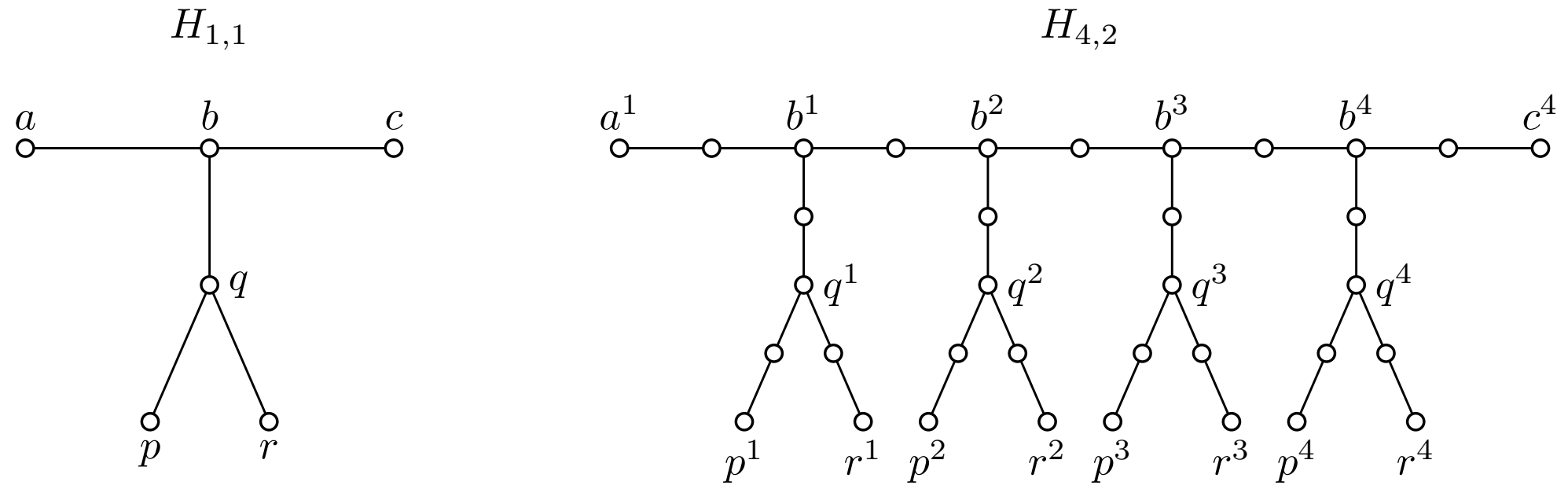
\begin{figure}[H]
    \centering
    \begin{tikzpicture}[
    x=1cm,y=1cm,
    vertex/.style={circle,draw=black,fill=white,inner sep=0pt,minimum size=4pt,line width=.65pt},
    edge/.style={draw=black,line width=.55pt},
    every node/.style={font=\small,inner sep=2pt},
]
    \foreach \copies/\subdivision/\panelx in {1/1/0,4/2/5.0} {
        \begin{scope}[xshift=\panelx cm]
            \pgfmathtruncatemacro{\lastcolumn}{\copies+1}
            \node at ({.775*\lastcolumn},1.0) {$H_{\copies,\subdivision}$};

            % The spine is a^1, b^1, ..., b^h, c^h.
            % At a shared vertex, b^i = a^{i+1} and c^i = b^{i+1}.
            \draw[edge] (0,0) -- ({1.55*\lastcolumn},0);
            \foreach \copyindex in {1,...,\copies} {
                \pgfmathsetmacro{\stemx}{1.55*\copyindex}
                \draw[edge] (\stemx,0) -- (\stemx,-1.15)
                    -- ({\stemx-.5},-2.3);
                \draw[edge] (\stemx,-1.15) -- ({\stemx+.5},-2.3);
            }

            % Draw the original vertices and their labels above the edges.
            \ifnum\copies=1\relax
                \node[vertex,label={above:$a$}] at (0,0) {};
                \node[vertex,label={above:$c$}] at ({1.55*\lastcolumn},0) {};
            \else
                \node[vertex,label={above:$a^1$}] at (0,0) {};
                \node[vertex,label={above:$c^{\copies}$}] at ({1.55*\lastcolumn},0) {};
            \fi
            \foreach \copyindex in {1,...,\copies} {
                \pgfmathsetmacro{\stemx}{1.55*\copyindex}
                \ifnum\copies=1\relax
                    \def\copysuperscript{}
                \else
                    \def\copysuperscript{^{\copyindex}}
                \fi
                \node[vertex,label={above:$b\copysuperscript$}]
                    at (\stemx,0) {};
                \node[vertex,label={right:$q\copysuperscript$}]
                    at (\stemx,-1.15) {};
                \node[vertex,label={below:$p\copysuperscript$}]
                    at ({\stemx-.5},-2.3) {};
                \node[vertex,label={below:$r\copysuperscript$}]
                    at ({\stemx+.5},-2.3) {};
            }

            % In H_{4,2}, each of the 17 edges of H_{4,1} has one midpoint.
            \ifnum\subdivision=2\relax
                \foreach \column in {1,...,\lastcolumn}
                    \node[vertex] at ({1.55*(\column-.5)},0) {};
                \foreach \copyindex in {1,...,\copies} {
                    \pgfmathsetmacro{\stemx}{1.55*\copyindex}
                    \node[vertex] at (\stemx,-.575) {};
                    \node[vertex] at ({\stemx-.25},-1.725) {};
                    \node[vertex] at ({\stemx+.25},-1.725) {};
                }
            \fi
        \end{scope}
    }
\end{tikzpicture}
    \caption{The trees $H_{1,1}$ (left) and $H_{4,2}$ (right).
    In the right panel, $b^i=a^{i+1}$ and $c^i=b^{i+1}$ for $i\in\{1,2,3\}$;
    unlabelled vertices subdivide each edge of $H_{4,1}$ once.}
    \label{fig:dbs-trees}
\end{figure}

For an integer $h \geq 1$, let $H_{h,1}$ be the tree created
from $h$ copies of $H_{1,1}$, denoted $H^1, \ldots,H^h$
(with vertices in the $i$-th copy denoted $a^i$, $b^i$, etc.),
where for $1 \leq i < h$ we identify $b^i$ with $a^{i+1}$
and $c^i$ with $b^{i+1}$ (this also identifies the edges
$b^ic^i$ and $a^{i+1}b^{i+1}$). 
Finally, for an integer $t \geq 1$, let $H_{h,t}$ be the tree
created from $H_{h,1}$ by replacing every edge
with a path of length $t$ (i.e., subdividing every edge $t-1$ times).
The trees $H_{1,1}$ and $H_{4,2}$ are shown in \cref{fig:dbs-trees}.

It is straightforward to see that every heeled caterpillar is an induced minor of $H_{h,t}$ for some $h$ and $t$. Thus, in order to obtain \Cref{thm:dbscaterpillar}, it is sufficient to prove the following result.

\begin{theorem}\label{thm:dbs-main}
For all integers $h, t \geq 1$, the class
of $H_{h,t}$-induced-minor-free graphs admits $h(9t+8)$-dominated balanced
separators.
\end{theorem}

\subsection{Preliminaries regarding dominated balanced separators and Gy\'arf\'as paths}

\paragraph{Dominated balanced separators.}
Let $w\colon V(G)\to \mathbb{R}_{\geq 0}$ be a weight function on vertices of a graph. By $w(A)$ we denote $\sum_{v\in A} w(v)$.
A set $A \subseteq V(G)$ is \emph{heavy} if $\omega(A) > \omega(V(G))/2$.
As $\omega$ is nonnegative, two disjoint sets cannot be both heavy. 
Consequently, for every set $Y \subseteq V(G)$, at most one connected component of $G-N[Y]$ has a heavy vertex set;
we call it the \emph{heavy component} of $G-N[Y]$, and we denote its vertex set by $D(Y)$.
If $G-N[Y]$ has no heavy component, then we set $D(Y) \coloneqq \emptyset$.

We say that set $S$ is a balanced separator of $G$ if for every connected component $C$ of $G-S$ we have $w(C)\leq \tfrac12 w(V(G))$.
Note that $N[S]$ is a balanced separator if and only if $D(S)=\emptyset$.
We say that $G$ admits \emph{$k$-dominated balanced separators} if there exists a balanced separator $S$ for $G$ and the set $X$ of size at most $k$ such that $S\subseteq N[X]$.

\paragraph{The cost of a set.}
For sets $X,A \subseteq V(G)$, let
\[
    \cost_X(A) \coloneqq \min\{|Y| : Y \subseteq V(G),\ D(X \cup Y) \cap A = \emptyset\}.
\]
This is well defined, as $D(V(G)) = \emptyset$.
A set $Y \subseteq V(G)$ with $|Y| = \cost_X(A)$ and $D(X \cup Y) \cap A = \emptyset$ is called a \emph{witness} for $\cost_X(A)$.
Thus, $\cost_X(A)$ measures the number of additional closed neighborhoods needed to make the heavy component avoid $A$ after deleting $N[X]$.
In particular, if $|Y| < \cost_X(A)$, then the heavy component of $G-N[X \cup Y]$ exists and contains a vertex of $A$.
We write $\cost = \cost_{\emptyset}$ and put
\[
    \rho(X) \coloneqq \cost_X(V(G)) = \min\{|Y| : Y \subseteq V(G),\ D(X \cup Y) = \emptyset\}.
\]

We now collect basic properties of heavy components and of the functions $\cost_X$.

\begin{lemma}\label{lem:dbs-basic}
        Let $X,X',Y,Y',A,A',M \subseteq V(G)$ be sets of vertices. Then the following hold.
        \begin{enumerate}[(a),ref=(\alph*)]
            \item\label{item:dbs-basic-heavy}
            If $C \subseteq V(G) \setminus N[Y]$ is a heavy set such that $G[C]$ is connected, then $C \subseteq D(Y)$.
            In particular, if $Y \subseteq Y'$, then $D(Y') \subseteq D(Y)$.
            \item\label{item:dbs-basic-mono}
            If $A \subseteq A'$, then $\cost_X(A) \leq \cost_X(A')$.
            \item\label{item:dbs-basic-sub}
            $\cost_X(A \cup A') \leq \cost_X(A) + \cost_X(A')$.
            \item\label{item:dbs-basic-nbhd}
            $\cost_X(N[A]) \leq |A|$ and $\cost_X(N[A \cup A']) \leq \cost_X(N[A]) + |A'|$.
            \item\label{item:dbs-basic-meet}
            If $|Y| < \cost_X(A)$, then $D(X \cup Y) \cap A \neq \emptyset$.
            In particular, if $|Y| < \rho(X)$, then $D(X \cup Y) \neq \emptyset$.
            \item\label{item:dbs-basic-witness}
            If $Y$ is a witness for $\cost_X(N[A])$, then $D(X \cup Y) \subseteq D(X \cup A)$.
            In particular, if $\cost_X(N[A]) < \rho(X)$, then $D(X \cup A) \neq \emptyset$.
            \item\label{item:dbs-basic-base-mono}
            If $X \subseteq X'$, then $\cost_{X'}(A) \leq \cost_X(A)$.
            \item\label{item:dbs-basic-transfer}
            If $Z$ is a witness for $\cost_X(N[M])$, then $D(X \cup Z \cup Y') \subseteq D(X \cup M \cup Y')$.
            Consequently,
            \[
                \rho(X) \leq \cost_X(N[M]) + \rho(X \cup M)
                \quad\text{and}\quad
                \cost_X(N[M] \cup A) \leq \cost_X(N[M]) + \cost_{X \cup M}(A).
            \]
        \end{enumerate}
\end{lemma}
    \begin{proof}
        For~\ref{item:dbs-basic-heavy}, observe that the connected graph $G[C]$ is a subgraph of $G-N[Y]$,
        so $C$ is contained in the vertex set of a single connected component of $G-N[Y]$.
        The weight of this component is at least $\omega(C) > \omega(V(G))/2$,
        so it is the heavy component of $G-N[Y]$, and $C \subseteq D(Y)$.
        For the second statement, we may assume that $D(Y') \neq \emptyset$.
        Then $D(Y')$ is a heavy set that induces a connected subgraph and is disjoint from $N[Y'] \supseteq N[Y]$,
        so the first statement applies to $C = D(Y')$.

        For~\ref{item:dbs-basic-mono}, note that every set $Y''$ with $D(X \cup Y'') \cap A' = \emptyset$ also satisfies $D(X \cup Y'') \cap A = \emptyset$.

        For~\ref{item:dbs-basic-sub}, let $Y_1$ and $Y_2$ be witnesses for $\cost_X(A)$ and $\cost_X(A')$, respectively.
        By~\ref{item:dbs-basic-heavy}, $D(X \cup Y_1 \cup Y_2) \subseteq D(X \cup Y_1) \cap D(X \cup Y_2)$,
        so $D(X \cup Y_1 \cup Y_2)$ is disjoint from $A \cup A'$.
        Hence $\cost_X(A \cup A') \leq |Y_1 \cup Y_2| \leq \cost_X(A) + \cost_X(A')$.

        For~\ref{item:dbs-basic-nbhd}, the set $Y = A$ satisfies $D(X \cup A) \cap N[A] = \emptyset$, so $\cost_X(N[A]) \leq |A|$.
        As $N[A \cup A'] = N[A] \cup N[A']$, the second inequality follows from~\ref{item:dbs-basic-sub} and from the first inequality applied to $A'$.

        Statement~\ref{item:dbs-basic-meet} is just a reformulation of the definition of $\cost_X(A)$;
        for the second part, take $A = V(G)$.

        For~\ref{item:dbs-basic-witness}, the set $D(X \cup Y)$ is disjoint from $N[A]$ by the definition of a witness, and from $N[X]$ by definition.
        If $D(X \cup Y) \neq \emptyset$, then $D(X \cup Y) \subseteq D(X \cup A)$ follows from~\ref{item:dbs-basic-heavy} applied to $C = D(X \cup Y)$.
        If, moreover, $\cost_X(N[A]) < \rho(X)$, then $|Y| < \rho(X)$, so $D(X \cup Y) \neq \emptyset$ by~\ref{item:dbs-basic-meet}, and hence $D(X \cup A)$ is nonempty.

        For~\ref{item:dbs-basic-base-mono}, use $D(X' \cup Y) \subseteq D(X \cup Y)$ from~\ref{item:dbs-basic-heavy}.

        For~\ref{item:dbs-basic-transfer}, the set $D(X \cup Z \cup Y')$ is contained in $D(X \cup Z)$, which is disjoint from $N[M]$.
        Thus it is a connected heavy set (or empty) disjoint from $N[X \cup M \cup Y']$, and~\ref{item:dbs-basic-heavy} gives the inclusion.
        For the first inequality, take a witness $Y'$ for $\rho(X \cup M)$.
        Then $D(X \cup Z \cup Y') = \emptyset$, so $\rho(X) \leq |Z \cup Y'| \leq \cost_X(N[M]) + \rho(X \cup M)$.
        For the second inequality, take a witness $Y'$ for $\cost_{X \cup M}(A)$.
        The inclusion shows that $D(X \cup Z \cup Y')$ avoids $A$, and it also avoids $N[M]$ because it is contained in $D(X \cup Z)$.
        Hence $Z \cup Y'$ witnesses the required upper bound on $\cost_X(N[M] \cup A)$.
\end{proof}

Taking $X=\emptyset$ gives the corresponding properties of $\cost$.

\paragraph{Gy\'arf\'as paths}

Let $Y \subseteq V(G)$.
    A sequence $(x_0,x_1,\ldots,x_\ell)$ of $\ell+1 \geq 1$ vertices of $G$ is a \emph{Gy\'arf\'as path with respect to $Y$}
    if for every $h \in \{0,1,\ldots,\ell\}$ the following two conditions hold:
    \begin{enumerate}[label=(G\arabic*),ref=(G\arabic*)]
        \item\label{item:dbs-G-reach}
        $x_h \in N[D(Y \cup \{x_0,\ldots,x_{h-1}\})]$;
        \item\label{item:dbs-G-next}
        if $h \geq 1$, then $x_h$ is adjacent to $x_{h-1}$ and $x_h \in D(Y \cup \{x_0,\ldots,x_{h-2}\})$.
    \end{enumerate}
    Here, we use the convention that $\{x_0,\ldots,x_{h-1}\} = \emptyset$ if $h = 0$, and $\{x_0,\ldots,x_{h-2}\} = \emptyset$ if $h = 1$.
    In words, every next vertex $x_h$ is a neighbor of the current last vertex $x_{h-1}$
    that lies in the heavy component after deleting the closed neighborhoods of $Y$ and of all vertices before $x_{h-1}$,
    and that is adjacent to the heavy component after deleting additionally the closed neighborhood of $x_{h-1}$.
    Clearly, every prefix of a Gy\'arf\'as path with respect to $Y$ is a Gy\'arf\'as path with respect to $Y$,
    and a single vertex $x_0$ forms a Gy\'arf\'as path $(x_0)$ with respect to $Y$ if and only if $x_0 \in N[D(Y)]$.

    \begin{lemma}\label{lem:dbs-gyarfas}
        Let $Y \subseteq V(G)$ and let $(x_0,x_1,\ldots,x_\ell)$ be a Gy\'arf\'as path with respect to $Y$.
        Then the following hold.
        \begin{enumerate}[(a),ref=(\alph*)]
            \item\label{item:dbs-gyarfas-induced}
            $x_0x_1\cdots x_\ell$ is an induced path in $G$, and $x_1,\ldots,x_\ell \in D(Y)$.
            In particular, the set $\{x_1,\ldots,x_\ell\}$ is disjoint from and anticomplete to $Y$.
            \item\label{item:dbs-gyarfas-last}
            If $\ell \geq 1$, then $x_\ell$ has a neighbor in $D(Y \cup \{x_0,\ldots,x_{\ell-1}\})$.
            \item\label{item:dbs-gyarfas-extend}
            If $D(Y \cup \{x_0,\ldots,x_\ell\}) \neq \emptyset$, then there exists a vertex $x_{\ell+1} \in V(G)$
            such that $(x_0,x_1,\ldots,x_\ell,x_{\ell+1})$ is a Gy\'arf\'as path with respect to $Y$.
        \end{enumerate}
    \end{lemma}
    \begin{proof}
        For~\ref{item:dbs-gyarfas-induced}, consider $h \in \{1,\ldots,\ell\}$.
        By~\ref{item:dbs-G-next}, $x_h$ is adjacent to $x_{h-1}$ (in particular, $x_h \neq x_{h-1}$) and $x_h \in D(Y \cup \{x_0,\ldots,x_{h-2}\})$.
        The latter set is disjoint from $N[x_g]$ for every $g \in \{0,\ldots,h-2\}$,
        so $x_h$ is different from and nonadjacent to each of $x_0,\ldots,x_{h-2}$.
        Hence, the vertices $x_0,\ldots,x_\ell$ are pairwise distinct and the only edges of $G$ between them are the edges $x_{h-1}x_h$ for $h \in \{1,\ldots,\ell\}$;
        that is, $x_0x_1\cdots x_\ell$ is an induced path.
        Moreover, $x_h \in D(Y \cup \{x_0,\ldots,x_{h-2}\}) \subseteq D(Y)$ by \cref{lem:dbs-basic}\ref{item:dbs-basic-heavy}.

        For~\ref{item:dbs-gyarfas-last}, let $Y_1 \coloneqq Y \cup \{x_0,\ldots,x_{\ell-1}\}$.
        By~\ref{item:dbs-G-reach} for $h = \ell$, we have $x_\ell \in N[D(Y_1)]$.
        Since $x_\ell$ is adjacent to $x_{\ell-1} \in Y_1$, we have $x_\ell \in N[Y_1]$, and hence $x_\ell \notin D(Y_1)$.
        Therefore, $x_\ell$ has a neighbor in $D(Y_1)$.

        For~\ref{item:dbs-gyarfas-extend}, let again $Y_1 \coloneqq Y \cup \{x_0,\ldots,x_{\ell-1}\}$, and let $Y_2 \coloneqq Y_1 \cup \{x_\ell\}$.
        By assumption, $D(Y_2) \neq \emptyset$, and $D(Y_2) \subseteq D(Y_1)$ by \cref{lem:dbs-basic}\ref{item:dbs-basic-heavy}.
        By~\ref{item:dbs-G-reach} for $h = \ell$, the set $D(Y_1)$ contains a vertex of $N[x_\ell]$,
        and this vertex does not belong to $D(Y_2)$, as $D(Y_2) \cap N[Y_2] = \emptyset$.
        Hence, $D(Y_2)$ is a nonempty proper subset of $D(Y_1)$.
        As $G[D(Y_1)]$ is connected, there is an edge $x'x''$ of $G$ with $x' \in D(Y_1) \setminus D(Y_2)$ and $x'' \in D(Y_2)$.
        We claim that $x'$ is a neighbor of $x_\ell$.
        Since $x' \in D(Y_1)$, we have $x' \notin N[Y_1]$.
        If also $x' \notin N[x_\ell]$, then $x'$ is a vertex of $G-N[Y_2]$ that is adjacent to the vertex $x''$ of the connected component $G[D(Y_2)]$ of $G-N[Y_2]$;
        but then $x' \in D(Y_2)$, a contradiction.
        Hence $x' \in N[x_\ell]$, and $x' \neq x_\ell$ because $x_\ell \in Y_2$ has no neighbor in $D(Y_2)$.
        We set $x_{\ell+1} \coloneqq x'$.
        Then $x_{\ell+1}$ is a neighbor of $x_\ell$ that belongs to $D(Y_1) = D(Y \cup \{x_0,\ldots,x_{\ell-1}\})$,
        so~\ref{item:dbs-G-next} holds for $h = \ell+1$;
        and $x_{\ell+1}$ has the neighbor $x''$ in $D(Y_2) = D(Y \cup \{x_0,\ldots,x_\ell\})$,
        so~\ref{item:dbs-G-reach} holds for $h = \ell+1$.
        For $h \leq \ell$, both conditions hold by assumption.
    \end{proof}

    \begin{lemma}\label{lem:dbs-gyarfas-long}
        Let $Y \subseteq V(G)$, let $x_0 \in N[D(Y)]$, and let $\ell \geq 1$ be an integer such that
        $D(Y \cup Y') \neq \emptyset$ for every set $Y' \subseteq V(G)$ with $|Y'| \leq \ell$.
        Then there exist vertices $x_1,\ldots,x_\ell \in V(G)$ such that $(x_0,x_1,\ldots,x_\ell)$ is a Gy\'arf\'as path with respect to $Y$.
    \end{lemma}
    \begin{proof}
        As $x_0 \in N[D(Y)]$, the sequence $(x_0)$ is a Gy\'arf\'as path with respect to $Y$.
        We extend it $\ell$ times using \cref{lem:dbs-gyarfas}\ref{item:dbs-gyarfas-extend}.
        This is possible: extending a Gy\'arf\'as path $(x_0,\ldots,x_h)$ with $h < \ell$,
        generates the set $Y' = \{x_0,\ldots,x_h\}$ with at most $h+1 \leq \ell$ elements, so $D(Y \cup Y') \neq \emptyset$ by assumption.
    \end{proof}

\subsection{Proof of \Cref{thm:dbs-main}}\label{sec:dbs-trees-main}

The key technical step in the proof of \cref{thm:dbs-main} is encapsulated in the following lemma.
\begin{lemma}\label{lem:dbs-step}
    Let $t \geq 1$ be an integer, 
    let $G$ be a connected graph with a weight function 
    $\omega : V(G) \to \mathbb{R}_{\geq 0}$, and let $v \in V(G)$.
    Then, there exists either a balanced separator of $(G,\omega)$
    consisting of at most $9t+8$ closed neighborhoods or
    an induced minor model $\mathcal{M}$ of $H_{1,t}$, two distinguished vertices $w, u \in V(G)$,
    and a set $X \subseteq V(G)$ of size at most $9t+8$
    with the following properties:
    \begin{enumerate}
    \item $D(X)$ is disjoint from and anticomplete to $\mathcal{M}$;
    \item in $\mathcal{M}$, $v$ belongs to the branch set of $a$
    and $w$ belongs to the branch set of $c$;
    \item $u \in N(D(X)) \cap N(w)$ and, furthermore, $w$ is the only neighbor of $u$ in $\mathcal{M}$.
    \end{enumerate}
\end{lemma}

Let us argue that \cref{lem:dbs-step} implies \cref{thm:dbs-main}.
\begin{proof}[Proof of~\cref{thm:dbs-main} assuming \cref{lem:dbs-step}.]
    Fix integers $h,t \geq 1$, and let $\beta \coloneqq 9t+8$.
    We show that for every $H_{h,t}$-induced-minor-free graph $G$ and every weight function
    $\omega \colon V(G) \to \mathbb{R}_{\geq 0}$,
    there is a set $Y \subseteq V(G)$ with $|Y| \leq h\beta$ such that $N[Y]$ is a balanced separator of $(G,\omega)$.

    It suffices to prove this for connected graphs, as otherwise we can simply focus on $D(\emptyset)$. Any balanced separator for this component is also a separator for the entire graph.

    Hence, assume that $G$ is connected, and put $W \coloneqq \omega(V(G))$.
    If $W=0$, the empty set is a balanced separator, so assume that $W>0$.
    We will prove the theorem by induction, repeatedly applying Lemma~\ref{lem:dbs-step}.
    After $i$ applications, we will either have found the required separator or have constructed
    a set $Y_i \subseteq V(G)$, an induced minor model $\mathcal{M}_i$ of $H_{i,t}$ in $G$,
    and vertices $w_i,u_i \in V(G)$ with the following properties:
    \begin{enumerate}[(a),ref=(\alph*)]
        \item $|Y_i| \leq i\beta$, and $G-N[Y_i]$ has a heavy component with vertex set $D_i$;
        \item $D_i$ is disjoint from and anticomplete to $V(\mathcal{M}_i)$;
        \item $w_i \in \mathcal{M}_i(c^i)$, $u_i \notin V(\mathcal{M}_i) \cup D_i$,
        $u_i \in N(D_i)$, and $N(u_i) \cap V(\mathcal{M}_i)=\{w_i\}$.
    \end{enumerate}
    Here $c^i$ is the leaf $c$ in the last copy of $H_{1,1}$ used to define $H_{i,1}$.

    For $i=1$, apply Lemma~\ref{lem:dbs-step} to $(G,\omega)$ and any vertex $v \in V(G)$.
    Its separator outcome proves the assertion, and its model outcome gives the stated properties with $Y_1=X$.
    In particular, $u_1$ lies outside the model because it has a neighbor in $D_1$,
    and it lies outside $D_1$ because it has the neighbor $w_1$ in the model.

    Suppose now that these objects have been constructed for some $i<h$.
    The graph
    \[
        F_i \coloneqq G[D_i \cup \{u_i\}]
    \]
    is connected, since $G[D_i]$ is connected and $u_i$ has a neighbor in $D_i$.
    Moreover, $u_i \in N_G[Y_i]$:
    otherwise, its neighbor in $D_i$ would place it in the same component of $G-N_G[Y_i]$, contrary to $u_i \notin D_i$.
    Define a weight function on $F_i$ by
    \[
        \omega_i(x) \coloneqq
        \begin{cases}
            \omega(x) & \text{if }x\in D_i,\\
            W-\omega(D_i) & \text{if }x=u_i.
        \end{cases}
    \]
    Thus, the weight outside $D_i$ is placed at $u_i$, and $\omega_i(V(F_i))=W$.
    Apply Lemma~\ref{lem:dbs-step} to $(F_i,\omega_i)$ with the prescribed vertex $u_i$.

    Suppose first that the lemma returns a set $Z \subseteq V(F_i)$ with $|Z|\leq \beta$
    such that $N_{F_i}[Z]$ is a balanced separator of $(F_i,\omega_i)$.
    Then $N_G[Y_i\cup Z]$ is a balanced separator of $(G,\omega)$.
    To see this, let $K$ be a~connected component of $G-N_G[Y_i\cup Z]$.
    It is contained in a component of $G-N_G[Y_i]$.
    If this component is not $G[D_i]$, then $\omega(V(K))\leq W/2$.
    Otherwise, $V(K)\subseteq D_i$, and $K$ is contained in a~component $K'$ of $F_i-N_{F_i}[Z]$.
    Since $\omega_i$ agrees with $\omega$ on $D_i$, we have
    \[
        \omega(V(K))=\omega_i(V(K))\leq\omega_i(V(K'))\leq W/2.
    \]
    Finally, $|Y_i\cup Z|\leq(i+1)\beta\leq h\beta$, as required.

    In the other outcome, let $\mathcal{N}$ be the induced minor model of $H_{1,t}$ returned by the lemma,
    let $Z$ be its set of at most $\beta$ vertices, let $D'$ be the vertex set of its heavy component,
    and let $w',u'$ be its distinguished vertices.
    Since $u_i\in\mathcal{N}(a)$ and $D'$ is disjoint from and anticomplete to $V(\mathcal{N})$,
    we have $D'\subseteq D_i$.
    Consequently, $\omega(D')=\omega_i(D')>W/2$.
    As $G[D']$ is a component of $F_i-N_{F_i}[Z]$ avoiding $u_i$,
    it is a component of $G[D_i]-N_G[Z]$.
    Since $G[D_i]$ is a component of $G-N_G[Y_i]$, it follows that $G[D']$ is also a component of $G-N_G[Y_i\cup Z]$.
    Set $Y_{i+1}\coloneqq Y_i\cup Z$ and $D_{i+1}\coloneqq D'$.
    Then $|Y_{i+1}|\leq(i+1)\beta$, and $G[D_{i+1}]$ is the unique heavy component after deleting $N_G[Y_{i+1}]$.

    We now join the two models.
    Their vertex sets are disjoint, and the only edge between them is $w_i u_i$:
    all vertices of the new model other than $u_i$ lie in $D_i$, which is anticomplete to the old model,
    and $w_i$ is the only neighbor of $u_i$ in the old model.
    Let $P_{ab}$ be the path of length $t$ between $a$ and $b$ in the new copy of $H_{1,t}$.
    Merge the branch sets
    \[
        \mathcal{M}_i(c^i)\ \cup\!
        \bigcup_{x\in V(P_{ab})}\mathcal{N}(x)
    \]
    into one branch set, and retain all other branch sets of the two models.
    The merged set is connected, since the edge $w_i u_i$ joins its two connected parts.
    This gives an induced minor model $\mathcal{M}_{i+1}$ of $H_{i+1,t}$.
    Indeed, the merged set models $c^i=b^{i+1}$, and the old path of length $t$ between $b^i$ and $c^i$
    now models the shared path between $a^{i+1}=b^i$ and $b^{i+1}=c^i$.
    The new paths between $b$ and $q$, between $q$ and $p$ and $r$, and between $b$ and $c$ are retained.
    The model conditions for $\mathcal{M}_i$ and $\mathcal{N}$, together with the fact that $w_i u_i$ is their only joining edge,
    ensure that no unwanted adjacencies arise.

    Set $w_{i+1}\coloneqq w'$ and $u_{i+1}\coloneqq u'$.
    The set $D_{i+1}\subseteq D_i$ is disjoint from and anticomplete to both models, hence also to $\mathcal{M}_{i+1}$.
    Furthermore, $u'$ lies outside $V(\mathcal{N})$ because it has a neighbor in $D'$,
    so $u'\neq u_i$ and therefore $u'\in D_i$.
    Thus, $u'$ is anticomplete to the old model, and its only neighbor in the combined model is $w'$,
    which belongs to the new terminal branch set of $c^{i+1}$.
    Since $u'$ has a neighbor in $D'$ and is adjacent to $w'$, all the required properties are preserved.

    After at most $h$ applications, this construction either returns a balanced separator consisting of at most $h\beta$ closed neighborhoods
    or produces an induced minor model of $H_{h,t}$ in $G$.
    The latter is excluded by assumption, so the required separator exists.
    As $h\beta=h(9t+8)\leq17ht$, this also gives the stated linear bound.
\end{proof}

We are left with proving \cref{lem:dbs-step}.
\begin{proof}[Proof of Lemma~\ref{lem:dbs-step}.]
    For an edge $xy$ of $H_{1,1}$, the $t-1$ inner vertices of the path that replaces $xy$ are called the \emph{subdivision vertices between $x$ and $y$} and are denoted by $xy_1, xy_2, \dots, xy_{t-1}$.
    We also set $xy_0=x$, $xy_t=y$, and $xy_i = yx_{t-i}$ for every $i\in\{0,1,\dots,t\}$.

    Let $s \coloneqq \cost(V(G))$.
    As $D(Y) \cap V(G) = \emptyset$ if and only if $D(Y) = \emptyset$, the integer $s$ is the minimum size of a set $Y \subseteq V(G)$ such that $N[Y]$ is a balanced separator of $(G,\omega)$.
    If $s \leq 9t+8$, then for a witness $Y$ for $\cost(V(G))$, the set $N[Y]$ is a balanced separator of $(G,\omega)$ consisting of at most $9t+8$ closed neighborhoods, and the lemma holds.
    Hence, from now on we assume that
    \begin{equation}\label{eq:dbs-s-large}
        s \geq 9t+9.
    \end{equation}
    In particular, $|\emptyset| = 0 < s = \cost(V(G))$, so $D(\emptyset) \neq \emptyset$ by the definition of $\cost$.
    As $G-N[\emptyset] = G$ is connected, we have $D(\emptyset) = V(G)$.

    Throughout the proof, we will be building an induced model $\mathcal{M}$ of $H_{1,t}$ in $G$.

    \medskip\noindent\textbf{Step 1: A long induced path starting at $v$.}
    Since $D(\emptyset) = V(G)$, the sequence $(v)$ is a Gy\'arf\'as path with respect to $\emptyset$.
    Starting from it, we apply \cref{lem:dbs-gyarfas}\ref{item:dbs-gyarfas-extend} as long as possible.
    As the vertices of a Gy\'arf\'as path are pairwise distinct by \cref{lem:dbs-gyarfas}\ref{item:dbs-gyarfas-induced}, this process stops,
    and we obtain a Gy\'arf\'as path $(v_0,v_1,\ldots,v_{n^\star})$ with respect to $\emptyset$ such that
    $v_0 = v$ and $D(\{v_0,\ldots,v_{n^\star}\}) = \emptyset$.
    For every $h \in \{0,\ldots,n^\star\}$, let
    \[
        \kappa_h \coloneqq \cost(N[\{v_0,v_1,\ldots,v_h\}]).
    \]
    By \cref{lem:dbs-basic}\ref{item:dbs-basic-nbhd}, we have $\kappa_0 \leq 1$,
    and by \cref{lem:dbs-basic}\ref{item:dbs-basic-mono} and~\ref{item:dbs-basic-nbhd},
    we have $\kappa_h \leq \kappa_{h+1} \leq \kappa_h + 1$ for every $h \in \{0,\ldots,n^\star-1\}$.
    Moreover, $\kappa_{n^\star} \geq s$ by \cref{lem:dbs-basic}\ref{item:dbs-basic-witness}, because $D(\{v_0,\ldots,v_{n^\star}\}) = \emptyset$.
    Hence, the sequence $\kappa_0,\kappa_1,\ldots,\kappa_{n^\star}$ is nondecreasing, increases by at most one in each step,
    starts with a value of at most $1$, and ends with a value of at least $s$;
    so it attains every integer value between $1$ and $s$.
    Since $1 < 4t+3 < 6t+4 < s$ by~\eqref{eq:dbs-s-large}, we can fix integers $m,n \in \{0,\ldots,n^\star\}$ with
    \[
        \kappa_m = 4t+3 \qquad\text{and}\qquad \kappa_n = 6t+4.
    \]
    As the sequence $(\kappa_h)_h$ is nondecreasing, we have $m < n$,
    and as it increases by at most one in each step, we have $n - m \geq \kappa_n - \kappa_m = 2t+1$.

    Let $P$ be the path $v_0v_1\dots v_n$; by \cref{lem:dbs-gyarfas}\ref{item:dbs-gyarfas-induced}, $P$ is an induced path in $G$.
    For integers $g,h \in \{0,\ldots,n\}$, let $P[g,h] \coloneqq \{v_g,v_{g+1},\ldots,v_h\}$;
    in particular, $P[g,h] = \emptyset$ if $g > h$.
    If $g \leq h$, then $G[P[g,h]]$ is the subpath $v_gv_{g+1}\cdots v_h$ of $P$.
    In the remainder of the proof, we will use the following properties of the path $P$:
    \begin{enumerate}[label=(P\arabic*),ref=(P\arabic*)]
        \item\label{item:dbs-P-induced} $P = v_0v_1\cdots v_n$ is an induced path in $G$ and $v_0 = v$;
        \item\label{item:dbs-P-m} $\cost(N[P[0,m]]) = 4t+3$;
        \item\label{item:dbs-P-n} $\cost(N[V(P)]) = 6t+4$; and
        \item\label{item:dbs-P-gap} $n - m \geq 2t+1$.
    \end{enumerate}
    Intuitively, by~\ref{item:dbs-P-m}, after deleting the closed neighborhoods of any $4t+2$ vertices,
    the heavy component still exists and contains a vertex of $N[P[0,m]]$;
    similarly, by~\ref{item:dbs-P-n}, after deleting the closed neighborhoods of any $6t+3$ vertices,
    the heavy component still exists and contains a vertex of $N[V(P)]$.
    Note that $4t+3 \leq |P[0,m]| = m+1$ by~\ref{item:dbs-P-m} and \cref{lem:dbs-basic}\ref{item:dbs-basic-nbhd},
    so $m \geq 4t+2$.
    Furthermore, $D(V(P)) \neq \emptyset$ by~\ref{item:dbs-P-n}, \eqref{eq:dbs-s-large}, and \cref{lem:dbs-basic}\ref{item:dbs-basic-witness}.

    \medskip\noindent\textbf{Step 2: The branch set of $q$, and the paths towards $p$ and $r$.}
    By~\ref{item:dbs-P-gap}, we have $m+t < n-t+1$,
    so $P[0,m+t]$ and $P[n-t+1,n]$ are the vertex sets of a prefix and a suffix of $P$ that are disjoint.
    We first find a vertex $z_q$ that is adjacent both to $P$ and to the heavy component of $G-N[V(P)]$,
    but has no neighbor in this prefix or in this suffix of $P$.

    \begin{claim}\label{clm:dbs-zq}
        There exists a vertex $z_q \in V(G) \setminus V(P)$ such that
        \begin{enumerate}[(a),ref=(\alph*)]
            \item\label{item:dbs-zq-P} $z_q$ has a neighbor in $V(P)$,
            \item\label{item:dbs-zq-D} $z_q$ has a neighbor in $D(V(P))$, and
            \item\label{item:dbs-zq-ends} $z_q$ has no neighbor in $P[0,m+t] \cup P[n-t+1,n]$.
        \end{enumerate}
    \end{claim}
    \begin{claimproof}
        Let $T \coloneqq P[0,m+t] \cup P[n-t+1,n]$.
        Hence, by \cref{lem:dbs-basic}\ref{item:dbs-basic-nbhd} and~\ref{item:dbs-P-m}, we have $$\cost(N[T]) \leq (4t+3) + 2t = 6t+3.$$
        Let $Y \subseteq V(G)$ be a witness for $\cost(N[T])$.
        Since $|Y| \leq 6t+3 < \cost(N[V(P)])$,  $D(Y)$ contains a vertex of $N[V(P)]$ by \cref{lem:dbs-basic}\ref{item:dbs-basic-meet}.
        We have $D(Y) \subseteq D(T)$ by \cref{lem:dbs-basic}\ref{item:dbs-basic-witness},
        so the set $D(T)$ contains a vertex of $N[V(P)]$ as well.

        On the other hand, $D(V(P)) \neq \emptyset$, as observed at the end of Step~1,
        and $D(V(P)) \subseteq D(T)$ by \cref{lem:dbs-basic}\ref{item:dbs-basic-heavy}, because $T \subseteq V(P)$.
        Since $D(V(P))$ is disjoint from $N[V(P)]$, we get that $D(V(P))$ is a nonempty proper subset of $D(T)$.
        As $G[D(T)]$ is connected, there is an edge $z_qz'$ of $G$ with $z_q \in D(T) \setminus D(V(P))$ and $z' \in D(V(P))$.
        
        We claim that $z_q$ has the desired properties.
        First, $z_q$ has the neighbor $z'$ in $D(V(P))$, which proves~\ref{item:dbs-zq-D}.
        Second, $z' \notin N[V(P)]$, so $z'$ has no neighbor in $V(P)$; hence, $z_q \notin V(P)$.
        Third, $z_q$ is adjacent to a vertex of the connected component $G[D(V(P))]$ of $G-N[V(P)]$,
        but $z_q$ does not belong to this component;
        hence, $z_q$ is not a vertex of $G-N[V(P)]$, that is, $z_q \in N[V(P)]$.
        Together with $z_q \notin V(P)$, this proves~\ref{item:dbs-zq-P}.
        Finally, $z_q \in D(T)$ and $D(T) \cap N[T] = \emptyset$, which proves~\ref{item:dbs-zq-ends}.
    \end{claimproof}

    We fix a vertex $z_q$ as in \cref{clm:dbs-zq}.
    Let $i$ be the smallest and $j$ be the largest index $h \in \{0,\ldots,n\}$ such that $v_h$ is adjacent to $z_q$;
    these indices exist by \cref{clm:dbs-zq}\ref{item:dbs-zq-P}.
    By \cref{clm:dbs-zq}\ref{item:dbs-zq-ends}, we have
    \begin{equation}\label{eq:dbs-ij}
        m+t+1 \leq i \leq j \leq n-t.
    \end{equation}
    We define the following sets of vertices (see also \Cref{fig:minor_model_middle_step} for illustration).
    \begin{itemize}
        \item We set $\mathcal{M}(q) \coloneqq \{v_i,v_j,z_q\}$.
        The graph $G[\mathcal{M}(q)]$ is connected, because $z_q$ is adjacent to both $v_i$ and $v_j$ (possibly $v_i = v_j$).
        The vertices of $P$ strictly between $v_i$ and $v_j$, that is, the vertices of $P[i+1,j-1]$, will not be used in the model.

        \item $L_p \coloneqq P[j+1,j+t]$.
        By~\eqref{eq:dbs-ij}, we have $j+t \leq n$, so $L_p$ consists of $t$ vertices that induce the subpath $v_{j+1}v_{j+2}\cdots v_{j+t}$ of $P$.
        The vertices $v_{j+1},\ldots,v_{j+t-1}$  model the subdivision vertices between $q$ and $p$, and $v_{j+t}$ will model $p$: we set $\mathcal{M}(qp_l) = \{v_{j+l}\}$ for $1\leq l\leq t-1$ and $\mathcal{M}(p)=\{v_{j+t}\}$.
        
        \item $L_r \coloneqq \{r_1,\ldots,r_t\}$, where the vertices $r_1,\ldots,r_t$ are defined as follows.
        By \cref{clm:dbs-zq}\ref{item:dbs-zq-D}, we have $z_q \in N[D(V(P))]$.
        Moreover, for every set $Y' \subseteq V(G)$ with $|Y'| \leq t$,
        \cref{lem:dbs-basic}\ref{item:dbs-basic-nbhd} and~\ref{item:dbs-P-n} give
        $\cost(N[V(P) \cup Y']) \leq (6t+4) + t < s$, so $D(V(P) \cup Y') \neq \emptyset$ by \cref{lem:dbs-basic}\ref{item:dbs-basic-witness}.
        Hence, \cref{lem:dbs-gyarfas-long} applied to $Y = V(P)$, $x_0 = z_q$, and $\ell = t$
        yields vertices $r_1,\ldots,r_t \in V(G)$ such that $(z_q,r_1,\ldots,r_t)$ is a Gy\'arf\'as path with respect to $V(P)$.
        We set $\mathcal{M}(r)=\{r_t\}$ and $\mathcal{M}(qr_l) = \{r_l\}$ for $1\leq l\leq t-1$.

        \item $F \coloneqq \mathcal{M}(q) \cup L_p \cup L_r$.
    \end{itemize}

    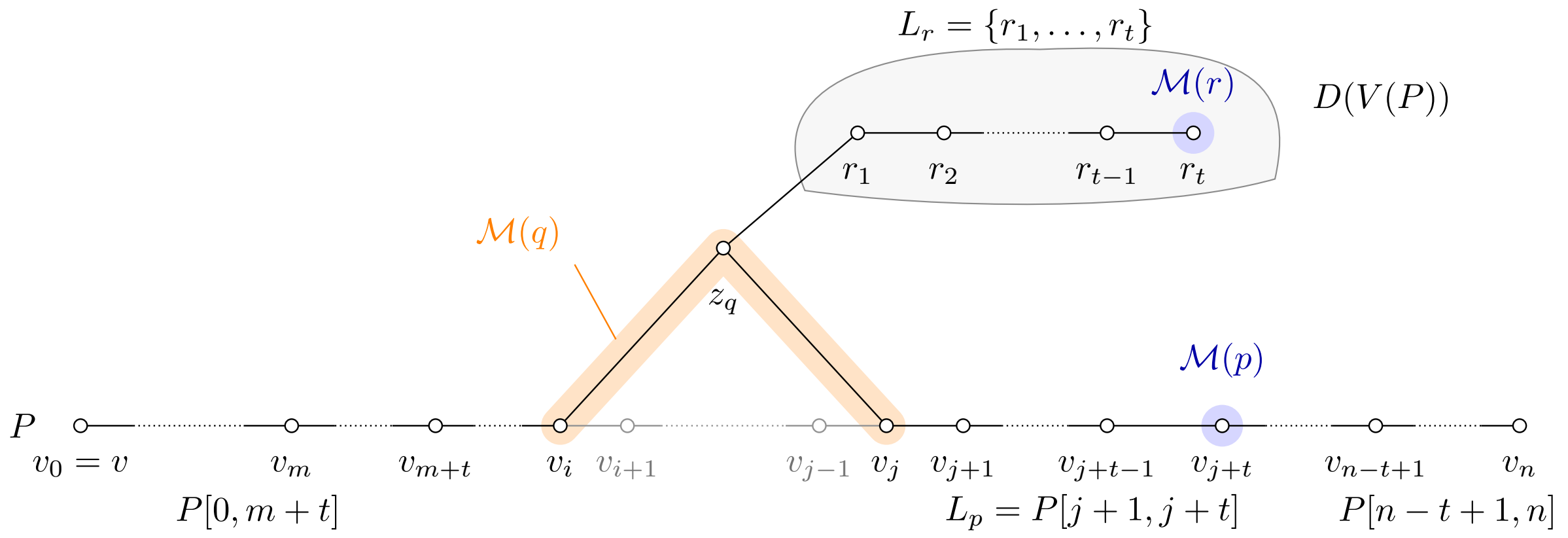
\begin{figure}[htbp]
    \centering
    \begin{tikzpicture}[x=.97cm,y=.97cm,
  vertex/.style={circle,draw=black,fill=white,inner sep=0pt,minimum size=3.8pt,line width=.45pt},
  unused/.style={vertex,draw=black!42},
  every node/.style={font=\small},
  edge/.style={draw=black,line width=.45pt},
  brace/.style={decorate,decoration={brace,amplitude=3pt},line width=.4pt}]
% A generic configuration with i<j. The path through the intervening vertices
% remains in G, but those vertices are omitted from the induced-minor model.
\coordinate (v0) at (0,0);
\coordinate (vm) at (2.2,0);
\coordinate (vmt) at (3.7,0);
\coordinate (vi) at (5.0,0);
\coordinate (vip) at (5.7,0);
\coordinate (vjm) at (7.7,0);
\coordinate (vj) at (8.4,0);
\coordinate (vjp) at (9.2,0);
\coordinate (vjtm) at (10.7,0);
\coordinate (vjt) at (11.9,0);
\coordinate (vnt) at (13.5,0);
\coordinate (vn) at (15.0,0);
\coordinate (zq) at (6.7,1.85);
\coordinate (r1) at (8.1,3.05);
\coordinate (r2) at (9.0,3.05);
\coordinate (rtm) at (10.7,3.05);
\coordinate (rt) at (11.6,3.05);
% M(q) is a narrow highlighted union along two edges, not the filled triangle:
% its interior P-vertices must not appear to be in the branch set.
\begin{scope}[on background layer]
  \path[fill=black!3,draw=black!45,line width=.4pt]
    (7.55,2.45) .. controls (7.0,3.75) and (8.8,3.95) .. (10.0,3.92)
    .. controls (12.0,4.0) and (12.7,3.7) .. (12.45,2.57)
    .. controls (11.2,2.22) and (8.9,2.26) .. cycle;
  \draw[orange!22,line width=11pt,line cap=round,line join=round]
    (vi)--(zq)--(vj);
  % Singleton branch sets of p and r.
  \fill[blue!16] (vjt) circle[radius=6pt];
  \fill[blue!16] (rt) circle[radius=6pt];
\end{scope}
% Original induced path P, with dots denoting unshown subpaths.
\draw[edge] (v0)--(.55,0);
\draw[edge,densely dotted] (.55,0)--(1.7,0);
\draw[edge] (1.7,0)--(vm)--(2.65,0);
\draw[edge,densely dotted] (2.65,0)--(3.25,0);
\draw[edge] (3.25,0)--(vmt)--(4.1,0);
\draw[edge,densely dotted] (4.1,0)--(4.6,0);
\draw[edge] (4.6,0)--(vi);
\draw[edge,black!42] (vi)--(vip)--(6.05,0);
\draw[edge,black!42,densely dotted] (6.05,0)--(7.35,0);
\draw[edge,black!42] (7.35,0)--(vjm)--(vj);
\draw[edge] (vj)--(vjp)--(9.6,0);
\draw[edge,densely dotted] (9.6,0)--(10.3,0);
\draw[edge] (10.3,0)--(vjtm)--(vjt)--(12.35,0);
\draw[edge,densely dotted] (12.35,0)--(13.05,0);
\draw[edge] (13.05,0)--(vnt)--(13.9,0);
\draw[edge,densely dotted] (13.9,0)--(14.6,0);
\draw[edge] (14.6,0)--(vn);
\draw[edge] (vi)--(zq)--(vj);
\draw[edge] (zq)--(r1)--(r2)--(9.4,3.05);
\draw[edge,densely dotted] (9.4,3.05)--(10.3,3.05);
\draw[edge] (10.3,3.05)--(rtm)--(rt);
\foreach \v in {v0,vm,vmt,vi,vj,vjp,vjtm,vjt,vnt,vn,zq,r1,r2,rtm,rt}
  \node[vertex] at (\v) {};
\foreach \v in {vip,vjm}\node[unused] at (\v) {};
% Vertex labels, separated from the braces underneath P.
\node[below=5pt] at (v0) {$v_0=v$};
\node[below=5pt] at (vm) {$v_m$};
\node[below=5pt] at (vmt) {$v_{m+t}$};
\node[below=5pt] at (vi) {$v_i$};
\node[below=5pt,text=black!55] at (vip) {$v_{i+1}$};
\node[below=5pt,text=black!55] at (vjm) {$v_{j-1}$};
\node[below=5pt] at (vj) {$v_j$};
\node[below=5pt] at (vjp) {$v_{j+1}$};
\node[below=5pt] at (vjtm) {$v_{j+t-1}$};
\node[below=5pt] at (vjt) {$v_{j+t}$};
\node[above=10pt,text=blue!65!black] at (vjt) {$\mathcal M(p)$};
\node[below=5pt] at (vnt) {$v_{n-t+1}$};
\node[below=5pt] at (vn) {$v_n$};
\node[below=8pt] at (zq) {$z_q$};
\node[below=5pt] at (r1) {$r_1$};
\node[below=5pt] at (r2) {$r_2$};
\node[below=5pt] at (rtm) {$r_{t-1}$};
\node[below=5pt] at (rt) {$r_t$};
\node[above=5pt,text=blue!65!black] at (rt) {$\mathcal M(r)$};
\node[left=9pt] at (v0) {$P$};
\node[anchor=west] at (12.7,3.4) {$D(V(P))$};
% Set labels. Orange is reserved for the three-vertex branch set.
\node[text=orange,anchor=east] (Mq) at (5.15,2.0)
  {$\mathcal M(q)$};
\draw[orange,line width=.45pt] (Mq.south east) -- (5.58,.9);
% Set labels retained without the annotated braces.
\node at (9.85,4.16) {$L_r=\{r_1,\ldots,r_t\}$};
\node at (10.55,-.9) {$L_p=P[j+1,j+t]$};
\node at (1.85,-.9) {$P[0,m+t]$};
\node at (14.25,-.9) {$P[n-t+1,n]$};
\end{tikzpicture}
    \caption{The path $P$ and already assigned branch sets of $\mathcal{M}$.}
    \label{fig:minor_model_middle_step}
    \end{figure}
    
    Note that for all vertices of $H_{1,t}$ for which we defined the value of $\mathcal{M}$ above, the assigned branch set induces a non-empty connected subgraph of $G$.
    Among these branch sets, $\mathcal{M}(q)$ is adjacent only to $\mathcal{M}(qp_1)$ and $\mathcal{M}(qr_1)$, $\mathcal{M}(p)$ only to $\mathcal{M}(qp_{t-1})$ and $\mathcal{M}(r)$ only to $\mathcal{M}(qr_{t-1})$.
    For every $1\leq l\leq t-1$ we have that $\mathcal{M}(qr_l)$ and $\mathcal{M}(qp_l)$ are adjacent only to $\mathcal{M}(qr_{l-1})$ and $\mathcal{M}(qr_{l+1})$; and $\mathcal{M}(qp_{l-1})$ and $\mathcal{M}(qp_{l+1})$ respectively.
    Therefore all adjacencies and non-adjacencies between all these branch sets are as desired.
    
    Let us summarize the properties of the vertices chosen above which will be used later:
    \begin{enumerate}[label=(Q\arabic*),ref=(Q\arabic*)]
        \item\label{item:dbs-Q-zq} $z_q \notin V(P)$, and every neighbor of $z_q$ in $V(P)$ belongs to $P[i,j]$
        (by \cref{clm:dbs-zq} and the choice of $i$ and $j$).
        In particular, $z_q$ has no neighbor in $P[0,i-1] \cup P[j+1,n]$.
        \item\label{item:dbs-Q-r} $z_qr_1r_2\cdots r_t$ is an induced path in $G$, and $L_r \subseteq D(V(P))$
        (by \cref{lem:dbs-gyarfas}\ref{item:dbs-gyarfas-induced}).
        In particular, $L_r$ is disjoint from and anticomplete to $V(P)$.
        \item\label{item:dbs-Q-F} $|F| \leq 3 + t + t = 2t+3$.
    \end{enumerate}

    \medskip\noindent\textbf{Step 3: The branch set of $b$.}
    The first $t$ vertices of $P$ will model $a$ and the subdivision vertices between $a$ and $b$: let us set $\mathcal{M}(a)=\{v_0\}$ and let $\mathcal{M}(ab_l)=\{v_l\}$ for $1\leq l\leq t-1$.
    The $t-1$ vertices of $P$ right before $v_i$ will model the subdivision vertices between $b$ and $q$: let us set $\mathcal{M}(qb_l)=\{v_{i-l}\}$ for $1\leq l\leq t-1$.

    All the branch sets we defined now are just single vertices, so they all induce non-empty connected subgraphs.
    As $P$ is induced, we get that the adjacencies between all branch sets contained in $V(P)$ are exactly as in $H_{1,t}$.
    By the choice of $i$ we have that $z_q$ has no neighbors inside $P[0,i-1]$, so $\mathcal{M}(q)$ is disjoint and anti-adjacent to the branch set of $a$ and branch sets of subdivision vertices between $a$ and $b$.
    The sets $\mathcal{M}(r)$ and $\mathcal{M}(qr_l)$ for $1\leq l\leq t-1$ are contained in $D(V(P))$, so these branch sets are disjoint and anti-adjacent to $V(P)$.
    Therefore model $\mathcal{M}$ defined so far is an induced minor model of the graph $H_{1,t}$ with $b, c$ and all subdivision vertices between them removed.
    
    Let
    \[
        S \coloneqq P[0,t-1] \cup P[i-t+1,i-1] \cup F.
    \]
    Observe that $S$ is the sum of all already defined branch sets of $\mathcal{M}$. We have
    $|S| \leq t + (t-1) + (2t+3) = 4t+2$ by~\ref{item:dbs-Q-F}.

    Since $m \geq 4t+2$, \eqref{eq:dbs-ij} implies $i-t \geq m+1 \geq 4t+3 > t$. The set $P[t,i-t]$ will be contained in the branch set of $b$.
    To complete this branch set, we will connect $P[t,i-t]$ to a heavy component that is anticomplete to $V(P)$ and to $F$,
    by a path that, apart from its first vertex, avoids the closed neighborhood of $S$.
    Let $Z \subseteq V(G)$ be a witness for $\cost(N[V(P)])$;
    that is, $|Z| = 6t+4$ by~\ref{item:dbs-P-n}, and $D(Z) \cap N[V(P)] = \emptyset$.
    Let
    \[
        \widehat{D} \coloneqq D(Z \cup F).
    \]

    \begin{claim}\label{clm:dbs-hatD}
        The following hold.
        \begin{enumerate}[(a),ref=(\alph*)]
            \item\label{item:dbs-hatD-avoid} $\widehat{D} \neq \emptyset$ and $\widehat{D} \cap N[V(P) \cup F] = \emptyset$;
            that is, $\widehat{D}$ is disjoint from and anticomplete to $V(P) \cup F$.
            \item\label{item:dbs-hatD-in-S} $\widehat{D} \subseteq D(S)$.
            \item\label{item:dbs-hatD-reach} $D(S)$ contains a vertex of $N[P[t,i-t]]$.
        \end{enumerate}
    \end{claim}
    \begin{claimproof}
        For~\ref{item:dbs-hatD-avoid}, we have $|Z \cup F| \leq (6t+4) + (2t+3) < s$ by~\ref{item:dbs-Q-F} and~\eqref{eq:dbs-s-large},
        so $\widehat{D} \neq \emptyset$ by \cref{lem:dbs-basic}\ref{item:dbs-basic-meet}.
        By \cref{lem:dbs-basic}\ref{item:dbs-basic-heavy}, we have $\widehat{D} \subseteq D(Z)$,
        and $D(Z) \cap N[V(P)] = \emptyset$ by the choice of $Z$.
        Furthermore, $\widehat{D} = D(Z \cup F)$ is disjoint from $N[F]$.

        For~\ref{item:dbs-hatD-in-S}, note that $S \subseteq V(P) \cup F$, so $\widehat{D} \cap N[S] = \emptyset$ by~\ref{item:dbs-hatD-avoid}.
        Since $\widehat{D}$ is heavy and induces a connected subgraph of $G$, \cref{lem:dbs-basic}\ref{item:dbs-basic-heavy} implies $\widehat{D} \subseteq D(S)$.

        For~\ref{item:dbs-hatD-reach}, note that $|S| \leq 4t+2 < 4t+3 = \cost(N[P[0,m]])$ by~\ref{item:dbs-P-m}.
        Hence, by \cref{lem:dbs-basic}\ref{item:dbs-basic-meet}, the set $D(S)$ contains a vertex of $N[P[0,m]]$.
        As $D(S) \cap N[S] = \emptyset$ and $P[0,t-1] \subseteq S$, this vertex does not belong to $N[P[0,t-1]]$.
        Consequently, it belongs to $N[P[t,m]] \subseteq N[P[t,i-t]]$, where the inclusion follows from $m \leq i-t$.
    \end{claimproof}

    The graphs $G[P[t,i-t]]$ and $G[D(S)]$ are connected,
    and by \cref{clm:dbs-hatD}\ref{item:dbs-hatD-reach}, they share a vertex or are joined by an edge.
    Hence, the graph $G[P[t,i-t] \cup D(S)]$ is connected.
    By \cref{clm:dbs-hatD}\ref{item:dbs-hatD-avoid} and~\ref{item:dbs-hatD-in-S}, it contains the nonempty set $\widehat{D}$,
    so it contains a path starting in $P[t,i-t]$ and ending in $\widehat{D}$.
    Let $y_0y_1\dots y_d$ be such a path of minimum possible length $d$ such that $y_0 \in P[t,i-t]$, $y_d \in \widehat{D}$.
    and $y_0,y_1,\ldots,y_d$ are vertices of $P[t,i-t] \cup D(S)$.

    \begin{claim}\label{clm:dbs-connector}
        The following hold.
        \begin{enumerate}[(a),ref=(\alph*)]
            \item\label{item:dbs-conn-S} $d \geq 2$ and $y_1,\ldots,y_d \in D(S) \setminus P[t,i-t]$.
            In particular, the set $\{y_1,\ldots,y_d\}$ is disjoint from and anticomplete to $S$.
            \item\label{item:dbs-conn-hatD} None of the vertices $y_0,\ldots,y_{d-1}$ belongs to $\widehat{D}$,
            none of the vertices $y_0,\ldots,y_{d-2}$ has a neighbor in $\widehat{D}$,
            and $y_{d-1}$ has a neighbor in $\widehat{D}$, namely $y_d$.
        \end{enumerate}
    \end{claim}
    \begin{claimproof}
        We use the minimality of $d$ several times.
        Since $\widehat{D} \cap V(P) = \emptyset$ by \cref{clm:dbs-hatD}\ref{item:dbs-hatD-avoid}, we have $y_d \notin P[t,i-t]$.
        If $y_h \in P[t,i-t]$ for some $h \in \{1,\ldots,d-1\}$, then $y_hy_{h+1}\cdots y_d$ would be a shorter path with the required properties.
        Hence, $y_1,\ldots,y_d \notin P[t,i-t]$, and thus $y_1,\ldots,y_d \in D(S)$.
        Similarly, if $y_h \in \widehat{D}$ for some $h \in \{0,\ldots,d-1\}$, then $y_0y_1\cdots y_h$ would be a shorter path with the required properties;
        so $y_0,\ldots,y_{d-1} \notin \widehat{D}$.
        If some $y_h$ with $h \in \{0,\ldots,d-2\}$ had a neighbor $y' \in \widehat{D}$,
        then $y_0y_1\cdots y_hy'$ would be a path (note that $y' \notin \{y_0,\ldots,y_h\}$ as $y' \in \widehat{D}$)
        with the required properties and of length $h+1 < d$, a contradiction.
        Finally, $y_0 \in V(P)$ neither belongs to $\widehat{D}$ nor has a neighbor in $\widehat{D}$, by \cref{clm:dbs-hatD}\ref{item:dbs-hatD-avoid};
        hence, $d \geq 2$.
        The remaining statements follow from $D(S) \cap N[S] = \emptyset$ and from $y_{d-1}y_d$ being an edge of $G$.
    \end{claimproof}

    We define the branch set of $b$ as
    \[
        \mathcal{M}(b) \coloneqq P[t,i-t] \cup \{y_1,\ldots,y_{d-1}\}.
    \]
    The graph $G[\mathcal{M}(b)]$ is connected: $P[t,i-t]$ induces a subpath of $P$,
    $y_1y_2\dots y_{d-1}$ is a path (with at least one vertex, as $d \geq 2$),
    and $y_1$ is adjacent to $y_0 \in P[t,i-t]$.
    Note that, in contrast to $\mathcal{M}(q)$, the branch set $\mathcal{M}(b)$ may contain many vertices of $P$.
    This is safe though,
    as $P$ is an induced path, the only vertices of $P$ outside $P[t,i-t]$ that are adjacent to $P[t,i-t]$ are $v_{t-1}$ and $v_{i-t+1}$,
    which are supposed to be adjacent to the branch set of $b$.
    By~\ref{item:dbs-Q-zq} and~\ref{item:dbs-Q-r}, $z_q$, $\mathcal{M}(r)$ and $\mathcal{M}(qr_l)$ for $1\leq l\leq t-1$ are anticomplete to $P[t,i-t]$.
    Therefore $\mathcal{M}(b)$ is adjacent exactly to the appropriate branch sets of vertices in $H_{1,t}$: $\mathcal{M}(ab_{t-1})$ and $\mathcal{M}(bq_1)$, and disjoint from and anti-adjacent to all others.

    \medskip\noindent\textbf{Step 4: The path towards $c$, and the vertex $u$.}
    By \cref{clm:dbs-connector}\ref{item:dbs-conn-hatD}, we have $y_{d-1} \in N[\widehat{D}] = N[D(Z \cup F)]$.
    Moreover, for every set $Y' \subseteq V(G)$ with $|Y'| \leq t+1$,
    we have $|Z \cup F \cup Y'| \leq (6t+4) + (2t+3) + (t+1) = 9t+8 < s$ by~\ref{item:dbs-Q-F} and~\eqref{eq:dbs-s-large},
    so $D(Z \cup F \cup Y') \neq \emptyset$ by \cref{lem:dbs-basic}\ref{item:dbs-basic-meet}.
    Hence, \cref{lem:dbs-gyarfas-long} applied to $Y = Z \cup F$, $x_0 = y_{d-1}$, and $\ell = t+1$
    yields vertices $c_1,\ldots,c_t,u \in V(G)$ such that $(y_{d-1},c_1,\ldots,c_t,u)$ is a Gy\'arf\'as path with respect to $Z \cup F$.
    We define
    \[
        L_c \coloneqq \{c_1,\ldots,c_t\}, \qquad
        w \coloneqq c_t, \qquad
        X \coloneqq Z \cup F \cup \{y_{d-1}\} \cup L_c, \qquad\text{and}\qquad
        D \coloneqq D(X).
    \]
    The vertices $c_1,\ldots,c_{t-1}$ will model the subdivision vertices between $b$ and $c$: let $\mathcal{M}(bc_l)=\{c_l\}$ for $1\leq l\leq t-1$ and let $\mathcal{M}(c) = \{c_t\}$.
    As all these sets are singletons, they clearly induce non-empty connected subgraphs.
    Note that $|X| \leq (6t+4) + (2t+3) + 1 + t = 9t+8$.
    Hence, $D \neq \emptyset$ by~\eqref{eq:dbs-s-large} and \cref{lem:dbs-basic}\ref{item:dbs-basic-meet},
    so $D$ is the vertex set of the unique connected component of $G-N[X]$ of weight more than $\omega(V(G))/2$.
    By \cref{lem:dbs-gyarfas}\ref{item:dbs-gyarfas-induced} and~\ref{item:dbs-gyarfas-last}, the following hold.
    \begin{enumerate}[label=(C\arabic*),ref=(C\arabic*)]
        \item\label{item:dbs-C-induced} $y_{d-1}c_1c_2\cdots c_tu$ is an induced path in $G$.
        \item\label{item:dbs-C-hatD} $c_1,\ldots,c_t,u \in \widehat{D}$.
        In particular, by \cref{clm:dbs-hatD}\ref{item:dbs-hatD-avoid}, $L_c \cup \{u\}$ is disjoint from and anticomplete to $V(P) \cup F$.
        \item\label{item:dbs-C-u} $u$ has a neighbor in $D(Z \cup F \cup \{y_{d-1},c_1,\ldots,c_t\}) = D$.
    \end{enumerate}

    By \ref{item:dbs-C-hatD} we get that the branch sets of $c$ and subdivision vertices of $b$ and $c$ are disjoint and anti-adjacent to all previously defined branch sets but $\mathcal{M}(b)$.
    By \ref{item:dbs-C-induced} we get that $\mathcal{M}(b)$ is connected to $\mathcal{M}(bc_1)$, $\mathcal{M}(c)$ is connected to $\mathcal{M}(bc_{t-1})$ and for every $1\leq l\leq t-1$ we have $\mathcal{M}(bc_l)$ is connected to $\mathcal{M}(bc_{l-1})$ and $\mathcal{M}(bc_{l+1})$.
    There are no other adjacencies by \ref{item:dbs-C-induced}, \cref{clm:dbs-connector}\ref{item:dbs-conn-hatD} and \ref{item:dbs-C-hatD}.

    Therefore $\mathcal{M}$ is now an induced model of $H_{1,t}$ in $G$.

    \medskip\noindent\textbf{Step 5: The remaining properties.}
    We have already shown that $|X| \leq 9t+8$ and that $D = D(X)$ is the vertex set of the unique connected component of $G-N[X]$ of weight more than $\omega(V(G))/2$.
    We now verify the three properties from the statement of the lemma.

    For the first property, we show that $D \cap N[V(\mathcal{M})] = \emptyset$.
    Since $Z \cup F \subseteq X$, \cref{lem:dbs-basic}\ref{item:dbs-basic-heavy} gives $D \subseteq \widehat{D}$,
    and hence $D \cap N[V(P) \cup F] = \emptyset$ by \cref{clm:dbs-hatD}\ref{item:dbs-hatD-avoid}.
    By \cref{clm:dbs-connector}\ref{item:dbs-conn-hatD}, none of the vertices $y_1,\ldots,y_{d-2}$ belongs to $\widehat{D}$ or has a neighbor in $\widehat{D}$,
    so $D \cap N[\{y_1,\ldots,y_{d-2}\}] = \emptyset$.
    Finally, $D \cap N[\{y_{d-1}\} \cup L_c] = \emptyset$, because $\{y_{d-1}\} \cup L_c \subseteq X$.

    The second property holds as $v = v_0 \in \mathcal{M}(a)$ and $w = c_t \in \mathcal{M}(c)$.

    For the third property, observe that $u$ is adjacent to $c_t = w$ by~\ref{item:dbs-C-induced}.
    In particular, $u \in N[X]$, so $u \notin D$.
    As $u$ has a neighbor in $D$ by~\ref{item:dbs-C-u}, we obtain $u \in N(D) \cap N(w)$.
    It remains to show that $u \notin V(\mathcal{M})$ and that $w$ is the only neighbor of $u$ in $V(\mathcal{M})$.
    By~\ref{item:dbs-C-hatD}, $u \in \widehat{D}$.
    Hence, $u \notin N[V(P) \cup F]$ by \cref{clm:dbs-hatD}\ref{item:dbs-hatD-avoid},
    and $u \notin N[\{y_1,\ldots,y_{d-2}\}]$ by \cref{clm:dbs-connector}\ref{item:dbs-conn-hatD}.
    Finally, by~\ref{item:dbs-C-induced}, $u \notin \{y_{d-1},c_1,\ldots,c_t\}$, and the only neighbor of $u$ among $y_{d-1},c_1,\ldots,c_t$ is $c_t = w$.
    Thus, $u \notin V(\mathcal{M})$ and $N(u) \cap V(\mathcal{M}) = \{w\}$.
    This finishes the proof of the lemma.
\end{proof}

\section{Conclusion}\label{sec:conclusion}
\Cref{conj:algorithmic} offered an appealing explanation of the tractability of \MWIS and \lcol{3} in many hereditary classes: excluding a planar graph as an induced minor was supposed to be the reason.
By the result of Bonnet and Chang~\cite{BonnetChang2026} and by our results, we now know that this explanation is not correct, even if polynomial time is relaxed to quasipolynomial time.
Nevertheless, \cref{conj:algorithmic} has been confirmed for a number of specific graphs $H$, including short paths~\cite{DBLP:conf/soda/LokshantovVV14,grzesik2020polynomialtime} and several further small graphs~\cite{dallard2022secondpaper,DBLP:journals/algorithmica/BonnetDGTW26},
and its quasipolynomial version is known to hold for all paths and cycles~\cite{DBLP:conf/focs/GartlandL20,DBLP:conf/sosa/PilipczukPR21,DBLP:conf/stoc/GartlandLPPR21}.
This leaves a wide gap between the graphs $H$ for which efficient algorithms are known and those for which we know hardness, and the main question left open by our work is to close it.

\begin{question}\label{q:classification}
    For which graphs $H$ are \MWIS and \lcol{3} solvable in polynomial, or quasipolynomial, time in $H$-induced-minor-free graphs?
\end{question}

Already the case of trees seems challenging.
On the one hand, quasipolynomial-time algorithms for \MWIS are known when $H$ is a path or a subdivided claw~\cite{DBLP:conf/focs/GartlandL20,DBLP:conf/sosa/PilipczukPR21,DBLP:conf/stoc/GartlandLMPPR24}, that is, a tree with at most one vertex of degree larger than two.
On the other hand, both problems are \NP-hard for each of the trees in \cref{fig:grid-trees}, which have four such vertices.
What happens for trees with two or three vertices of degree larger than two?

\paragraph{Independent sets versus colorings.}
In all cases resolved in this paper, \MWIS and \lcol{3} behave in the same way:
our hardness results for both problems are obtained in the same host families $\cB$ and $\cT$, although for $\cT$ the coloring reduction requires passing to the induced minor closure.
This raises the following question.

\begin{question}\label{q:agree}
    Is it true that for every graph $H$, \MWIS is polynomial-time (resp.\ quasipolynomial-time) solvable in $H$-induced-minor-free graphs if and only if \lcol{3} is?
\end{question}

Note that no such correspondence holds for classes defined by excluding an induced subgraph.
For example, \MWIS is polynomial-time solvable in claw-free graphs~\cite{MINTY1980284,SBIHI198053}, while \col{3} is \NP-hard already in line graphs~\cite{Holyer1981}, which are claw-free.
A positive answer to \cref{q:agree} would therefore reflect a specific feature of induced minors.
The most natural approach would be a reduction from one problem to the other that preserves excluding a fixed induced minor.
For instance, the standard reduction from \lcol{3} to \MIS replaces every vertex $v$ with a clique whose vertices correspond to the colors in the list of $v$, and makes two copies of the same color adjacent whenever the corresponding vertices are adjacent.
In case of all list are just colors $\{1,2,3\}$, the resulting graph contains the input graph as an induced minor, but it may also contain other induced minors, and it is not clear whether it can be made to avoid a fixed planar graph $H$ whenever the input graph does.

\paragraph{Dominated balanced separators.}
Our results show that \cref{conj:dominated-separators}, even where it holds, does not by itself explain tractability:
for the tree $F_0$ in \cref{fig:grid-trees}, constantly dominated balanced separators coexist with the \NP-hardness of \MIS.
\Cref{conj:dominated-separators} itself remains open.
In particular, it is natural to ask 
whether \cref{conj:dominated-separators} holds for all trees $H$.
Finally, given that dominated balanced separators are not sufficient, which additional structural property of $H$-induced-minor-free graphs, for planar $H$, could govern the complexity of \MWIS and \lcol{3}?

\paragraph{Acknowledgement.} The authors are grateful to Maria Chudnovsky and Amadeus Reinald for stimulating discussions. 

\section*{AI usage statement}
During research and preparation of the manuscript, we have relied heavily on AI tools: GPT-6-Astra (via a grant of access to OpenAI models through the ChatGPT for Academic Researchers program) and Opus-5.5 (via Claude Team plan for research teams). 

\paragraph{Research.} 
The proofs in \cref{sec:grids,sec:color} are due to GPT-6-Astra.

\paragraph{Writing.} We have used AI tools in the following way while preparing the manuscript. %
\begin{itemize}
\item Opus 5.5 was used to polish the introduction.
\item The proofs in \cref{sec:grids,sec:NPmis} have been written by GPT-6.1-Sol according to the draft written by the authors, and then verified and polished by the authors.
\item The proofs in \cref{sec:color} have been written by GPT-6-Astra with a few rounds of proof simplification, notation changes, and verification by the authors.
    \item The proof of \Cref{thm:dbs-main} has been written down 
    first as a sketch by the authors, and then turned into the final write-up
    by a few rounds of ping-pong between GPT-6-Astra, Opus-5.5, and human verification and polishing.
\end{itemize}

\bibliographystyle{abbrv}
\bibliography{main}
\end{document}